\documentclass{sig-alternate} 
\usepackage{mathptmx} % This is Times font

\usepackage{fancyhdr}
\usepackage[hyphens]{url}
\usepackage{hyperref}
\usepackage{caption}

\newcommand{\ignore}[1]{}
\usepackage{fancyhdr}
\usepackage[normalem]{ulem}
\usepackage{array}
\usepackage[caption=false,font=footnotesize]{subfig}
\usepackage{fancyhdr}
\usepackage{mathptmx} % This is Times font
\usepackage{multirow}
\usepackage{amsmath}
\usepackage{bm}

\usepackage{pifont}

\usepackage{fancyhdr}
\usepackage[normalem]{ulem}
\usepackage[hyphens]{url}
\usepackage[sort,nocompress]{cite}
\usepackage[final]{microtype}
\usepackage{flushend}
\usepackage{hhline}
\usepackage{algorithm}
\usepackage{algorithmic}
\usepackage{hyperref}
\usepackage[hyphenbreaks]{breakurl}
\usepackage{url}

\newcommand\redout{\bgroup\markoverwith{\textcolor{red}{\rule[0.5ex]{2pt}{1pt}}}\ULon}

\usepackage{authblk}
\title{Tackling Variabilities in Autonomous Driving}
\vspace{-1cm}
\author[1, 2]{Yuqiong Qi}
\author[3]{Yang Hu}
\author[1]{Haibin Wu}
\author[5]{Shen Li}
\author[4]{Haiyu Mao}
\author[1]{Xiaochun Ye}
\author[1]{Dongrui Fan}
\author[1]{Ninghui Sun}
\affil[1]{SKLCA, Institute of Computing Technology, Chinese Academy of Sciences, Beijing, China}
\affil[2]{University of Chinese Academy of Sciences, Beijing, China}
\affil[3]{University of Texas at Dallas, USA}
\affil[4]{Tsinghua University, Beijing, China}
\affil[5]{National University of Singapore, Singapore}

\begin{document}
\maketitle
\thispagestyle{empty}
\pagestyle{plain}

%\iffalse
\vspace{-10em} 
\begin{abstract}
The state-of-the-art driving automation system demands extreme computational resources to meet rigorous accuracy and latency requirements. Though emerging driving automation computing platforms are based on ASIC to provide better performance and power guarantee, building such an accelerator-based computing platform for driving automation still present challenges. First, the workloads mix and performance requirements exposed  to  driving  automation  system  present  significant variability. Second, with more cameras/sensors integrated in a future fully autonomous driving vehicle, a heterogeneous multi-accelerator architecture substrate is needed that requires a design space exploration for a new form of parallelism. In this work, we aim to extensively explore the above system design challenges and these challenges motivate us to propose a comprehensive framework that synergistically handles the heterogeneous hardware accelerator design principles, system design criteria, and task scheduling mechanism. Specifically, we propose a novel heterogeneous multi-core AI accelerator (HMAI) to provide the hardware substrate for the driving automation tasks with variability. We also define system design criteria to better utilize hardware resources and  achieve increased throughput while satisfying the performance and energy restrictions. Finally, we propose a deep reinforcement learning (RL)-based task scheduling mechanism FlexAI, to resolve task mapping issue. Experimental results show that with FlexAI scheduling, basically 100\% tasks in each driving route can be processed by HMAI within their required period to ensure safety, and FlexAI can also maximally reduce the breaking distance up to 96\% as compared to typical heuristics and guided random-search-based algorithms.

\end{abstract}
\vspace{-1em} 
\section{Introduction}
The state-of-the-art driving automation system demands extreme computational resources to meet rigorous accuracy and latency requirements. The traditional single accelerator-based on-car computer system, such as Mobileye EyeQ3~\cite{EyeQ3}, and integrated GPU based platform, such as Nvidia Drive PX2, have shown staggering performance and limited processing capability when facing this performance demands ~\cite{2018-q3}. Therefore, emerging driving automation computing platforms such as Tesla FSD~\cite{FSD} and Horizon Journey 2~\cite{Robotics2019} propose ASIC-based computing platform that targets explicitly to neural-network-based task processing with regards to both performance and energy considerations. However, building such an accelerator-based computing platform for driving automation present following challenges.

First, the workloads mix and performance requirements exposed to driving automation system present significant \textit{variability}. For example, an automated driving vehicle usually operates a variety of concurrent neural network (NN)-based tasks, such as visual perception-based assistant driving (e.g., object detection and recognition, scene segmentation)~\cite{abs-1902-03589}, high-level route planning (e.g. localization based on HD map)~\cite{yang2019re}, HD map crowdsourcing~\cite{Slovick}, and intelligent human-machine interaction(e.g., voice recognition, gesture recognition, and driver eye-tracking). Furthermore, an automated vehicle typically equips multiple cameras and sensors where different cameras (e.g., forward-facing or side-facing) have differentiated stream generation rates and accuracy requirements under different driving areas (e.g., urban, undivided-highway or highway) and different driving behaviors (e.g., going straight, turning or reversing). \textit{Such variability shown in performance requirements and workload mix significantly challenges the performance guarantee and complicates the design of underlying computing platform in driving automation system}. 

Second, with the increasing number of cameras and camera frame rate in automated vehicles, driving automation system will generate enormous data for real-time analysis (e.g., 1200 frames per second (FPS)). Unfortunately, most single accelerator-based platforms cannot meet this overwhelming processing requirements. For example, ADM-7V3 FPGA\cite{ding2019req} only supports 314.2 FPS and Virtex-7 VC707\cite{nguyen2019high} only supports 109.3 FPS. With more cameras/sensors integrated in a future fully autonomous driving vehicle (L5) and increasing frame rate of cameras~\cite{Wang2019}, a multiple accelerator-based computing platform is desired. Considering the composite workload mix often involves running multiple NN models with distinct layer operations and sizes, this will call for a heterogeneous multi-accelerator architecture substrate that requires a design space exploration for a new form of parallelism. As advocated in ~\cite{abs-1907-02064}, the performance or efficiency of future computer systems will have to rely on new accelerator-level parallelism (ALP). A high ALP implies that each accelerator can execute a targeted computation class faster and usually with less energy.
%On the other hand, the processing workloads exposed to modern intelligent autonomous driving SoCs present increasing heterogeneity and further challenges the performance and energy of multiple accelerators. 

%Considering the composite workload mix often involves running multiple NN models with distinct layer operations and sizes. This will call for a multiple heterogeneous accelerator architecture that requires a new form of parallelism. As advocated in ~\cite{abs-1907-02064}, the performance or efficiency of future computer systems will have to rely on new accelerator-level parallelism (ALP). The ALP is defined as the parallelism of workload components (e.g. NNs for different cameras) concurrently executing on multiple accelerators. A high ALP implies that each accelerator can execute a targeted computation class faster and usually with less energy.

%To accommodate such many various tasks with differentiated accuracy and latency requirements, future multiple heterogeneous accelerators for autonomous driving must achieve a high ALP. This implies that allocating heterogeneous neural networks for cameras and sensors to multiple heterogeneous accelerators should consider both the restrictions of accuracy and camera-to-recognition latency, while the hardware resource should be maximally utilized and energy is minimally consumed. However, designing such a system is non-trivial.

These observations prompt us to consider an important question: \textbf{\textit{how can a driving automation computing system well-manage these challenging design aspects under rigorous performance and energy restrictions?}} To wit, to efficiently process such a large amount of CNN-based tasks with  high variability on the complicated heterogeneous hardware substrate, effective criteria for system design that are tailored to driving automation should be defined, and efficient task scheduling mechanism should be explored to meet the criteria. Unfortunately, current computing systems for driving automation have not provided answers to this question.  

In this work, we aim to extensively explore the above system design challenges and build a comprehensive driving automation framework that synergistically handles the key design aspects. \textit{First}, our framework features a novel heterogeneous multi-core AI accelerator (HMAI) to provide the hardware substrate for the driving automation tasks with variability. We also propose the design principle to choose the accelerators for the HMAI. 
%Specifically, we investigate heterogeneous object recognition or tracking tasks processed by three typical CNN-based programs YOLO\cite{yolo9000}, SSD\cite{ssd513}, and GOTURN\cite{GOTURN}, which are used towards cameras with different functions. Our rationale stands on the fact that vision processing tasks dominate the computational workloads in modern autonomous driving computing platforms~\cite{yang2019re}~\cite{abs-1902-03589}, while CNNs are essential building blocks for vision processing applications. We characterize YOLO, SSD and GOTURN on three representative CNN accelerators built based on our CNN accelerator taxonomy, and using these three accelerators to construct our HMAI. 
\textit{Second}, our framework defines system design criteria to better utilize hardware resources and achieve increased throughput while satisfying the performance and energy restrictions. Specifically, we propose two metrics, Matching Score (MS) and Global State Value (Gvalue) to formalize the criteria. MS pays attention to the safety requirements of various tasks in driving automation systems, while Gvalue puts more weight onto the overall performance of HMAI that reflects the globality.
%As we know, above mapping on HMAI is an NP-complete problem and is generally solved using heuristic\cite{11,ATA,H1,H2,H3,H4,H5,EDP} or guided random-search-based algorithms\cite{hou,shroff,correa,Haluk}. However, the scheduling strategies based on these algorithms  fail to see the global situation of HMAI like current resource utilization, the longest execution time among all cores, which often results in a suboptimal allocation.
\textit{Finally}, our framework employs a deep reinforcement learning (RL)-based task scheduling mechanism FlexAI, to resolve the task mapping issue. Specifically, we show that a robust policy can be yielded by applying deep Q-network (DQN)\cite{DQN}. The RL agent in FlexAI is predictive and global. The predictive means each policy will schedule a corresponding task immediately without considering the later-coming tasks. The global feature in FlexAI means it can consider the whole performance in hardware, such as resource utilization. 
In an autonomous driving system, each task should be processed within a certain period to ensure driving safety. For instance, if a static obstacle cannot be detected by a moving vehicle within a specified period, it may cause a traffic accident. In this paper, with FlexAI scheduling, basically 100\% tasks in each driving route can be processed by HMAI within their required period to ensure safety, while typical heuristics (e.g Min-Min), and guided random-search-based algorithms (GA, SA) can only ensure 21\%, 34\% and 51\% safety on average, respectively. Moreover, FlexAI can maximally reduce the breaking distance up to 96\% as compared to the above algorithms, thus efficiently avoiding traffic collision (traffic collision cause 36,096 people died in the United States in 2019~\cite{NHTSA}). In addition, our experimental results also show that FlexAI achieves up to 87\% execution time reduction and 960\% resource utilization balance rate (introduced in Section 6.2) improvement, outperforming typical heuristics, guided random-search-based algorithms, and unscheduled algorithms. Moreover, HMAI can achieve up to 5× speedup and 2.5× TOPS/W over NVIDIA Tesla T4 GPU, and 2.1× TOPS/W over homogeneous hardware platforms (detail in Section 3.1).
%In autonomous driving system, each task should be processed within a certain period to ensure driving safety. For instance, if a stationary obstacle cannot be detected within a specified period by a moving vehicle, it may lead to a traffic accident. With the scheduling of FlexAI, basically 100\% tasks in each driving route can be processed by HMAI within their required period to ensure safety, while typical heuristic (Min-Min), and guided random-search-based algorithms (GA, SA) can only ensure average 21\%, 34\% and 51\% tasks’ safety. Moreover, FlexAI can maximally reduce the breaking distance up to 96\% compared to above algorithms, thus efficiently avoid traffic collision (traffic collision cause 36,096 people died in the United States in 2019~\cite{NHTSA}). In addition, our experimental results also show that FlexAI achieves up to 87\% execution time reduction and 960\% resource utilization balance rate improvement with typical heuristic, guided random-search-based algorithms, and unscheduled algorithms. HMAI can achieve up to 5× speedup, and 2.5× TOPS/W over NVIDIA Tesla T4, and 2.1× TOPS/W over homogeneous hardware platforms.

\section{Variabilities in Driving Automation}

%Both industry\cite{Audi,Tesla,Waymo} and academia\cite{UCSan,Illinois,Texas,Safety} are intensively exploring the driving automation system.The Society of Automobile Engineers (SAE)\cite{driving2014levels} defines six levels of driving automation from Level 0 to Level 5, where Level 5 is fully autonomous. To deliver a functional and practical driving automation system with a safety and reliability guarantee, designers should carefully analyze and optimize every technical detail from hardware through software. In this section, we explore the system design challenges imposed by the variability of workloads and performance requirements in driving automation and the heterogeneity of multi-accelerators. These challenges motivate us to propose a comprehensive framework that synergistically handles the heterogeneous hardware accelerator design principles, system design criteria, and task scheduling mechanism. 

In this section, we show that the workload mix and performance requirements caused by the scenario change in driving automation are highly variable. 
%\textbf{The Need for Multiple Accelerators in Automated Vehicles.}

%\vspace{-0.7em} 
%\subsubsection{Heterogeneous CNN Models are Needed for Driving Automation Processing.}
%\vspace{-0.3em} 
\subsection{Heterogeneous CNN Models are Needed for Driving Automation Processing} 
The perception of an automated vehicle consists of a variety of workloads such as object detection (DET), object tracking (TRA), and localization (LOC) along the camera-to-recognition path. These diverse workloads can incur complex processing patterns to the underlying multi-accelerators. According to\cite{TAIAD}, DET, TRA, and LOC dominate the computing of the driving automation system. While for DET and TRA, the convolutional neural network (CNN)-based computation accounts for more than \textbf{94\%} of the execution time. Moreover, the DET and TRA are completely based on CNN-based camera data processing in Tesla\cite{jc}, and even the LIDAR-based DET algorithms are mostly based on CNN\cite{voxelnet}. Therefore, we will focus on CNN-based tasks in the driving automation system. Specifically, based on the research\cite{TAIAD}, we will focus on three typical CNN algorithms, the YOLO\cite{yolo9000} and SSD\cite{ssd513} for DET, and GOTURN\cite{GOTURN} for TRA. As shown in Table~\ref{amount}, we compare the key features of these three CNN algorithms.

\begin{table}[t]
\renewcommand\arraystretch{1}
  \centering
  \scriptsize
  %\begin{tabular}{ |p{1cm}<{\centering}|p{1cm}<{\centering} |p{1.4cm}<{\centering} |p{0.7cm}<{\centering} |p{2.2cm}<{\centering} |}
  \begin{tabular}{ p{0.8cm}<{\centering}p{1.1cm}<{\centering} p{2.8cm}<{\centering}p{1.3cm}<{\centering}}
    \hline
   \textbf{CNNs} & \textbf{$\sharp$of MACs } & \textbf{$\sharp$of weights and neurons}&\textbf{Layers num} \\
    \hline
    \textbf{SSD} & 26G& 697.76 M &53 \\
    \hline
    \textbf{YOLO}  & 16G &150M  &101\\
    \hline
    \textbf{GOTURN} &11G &13.95M &11\\
    \hline
    %AlexNet &55.96 M &3.87 G &1&3\\
    %\hline
     %VGG &246.95 G&31.44 T &0&1\\
    %\hline
  \end{tabular}
    \vspace{-1em} 
    \captionsetup{font={small}} 
    \caption{\textbf{The features of typical CNN algorithms in DET and TRA.}\label{amount}}
\end{table}

\begin{table}[t]
\renewcommand\arraystretch{1}
% \vspace{-0.8em} 
  \centering
  \scriptsize
  \begin{tabular}{ p{1.5cm}<{\centering}p{1.5cm}<{\centering} p{1.3cm}<{\centering} p{1cm}<{\centering} }
    \hline
\textbf{Object}& \textbf{Distance} & \textbf{Area} &\textbf{Proportion}\\ \hline
\multirow{2}{*}{\textbf{Vehicle}} & 163m & 4620  & 0.33\%  \\ \cline{2-4} 
                  &17.98m  & 42000 & 3\%  \\ \hline
\multirow{2}{*}{\textbf{Pedestrian}} &140m  & 4620 & 0.33\%   \\ \cline{2-4} 
                  & 15.48m & 42000 & 3\% \\ \hline
  \end{tabular}
    \vspace{-1em} 
    \captionsetup{font={small}} 
    \caption{\textbf{The area and proportion of the objects in the image when they are at different distances from the current vehicle.}\label{distance}}
\end{table}

\begin{table}[h]
\renewcommand\arraystretch{1}
  \centering
  \scriptsize
  \begin{tabular}{ p{1.5cm}<{\centering}p{1.5cm}<{\centering} p{0.8cm}<{\centering} p{0.8cm}<{\centering} p{0.8cm}<{\centering} }
    \hline
  \textbf{Method}& \textbf{Backbone} & \boldsymbol{$AP_S$} &\boldsymbol{$AP_M$}&\boldsymbol{$AP_L$} \\
    \hline
    \textbf{YOLOv2}\cite{yolo9000} & DarkNet-53&\textbf{18.3}&\textbf{35.4}&41.9 \\
    \hline
    \textbf{SSD312}\cite{ssd513}&ResNet-101&6.2&28.3&\textbf{49.3}\\
    \hline
    \textbf{SSD512}*\cite{ssd512}&VGG-16&10.9&31.8&\textbf{43.5}\\
    \hline
    \textbf{SSD513}\cite{ssd513}&ResNet-101&10.2&34.5&\textbf{49.8}\\
    \hline
  \end{tabular}
    \vspace{-1em} 
    \captionsetup{font={small}} 
    \caption{\textbf{Detection results of YOLO and SSD. $AP_S$ is AP of small objects, $AP_M$ is AP of medium objects and $AP_L$ is AP of large objects.}\label{AP}}
\end{table}

In this paper, we consider a co-run of multiple heterogeneous CNN-based perception models on a driving automation computing platform. This is motivated by the fact that driving automation scenarios present a diverse set of objects with different sizes, areas, and dynamic distances, which incur stringent and time-varying performance restrictions to the computing tasks. Take the representative driving automation data set KITTI\cite{KITTI} as an example. We show how the area of the representative object (vehicle and pedestrian) and its ratio to the area of entire image change with different distances in Table~\ref{distance}. Here the area is measured as the number of pixels in the segmentation mask. The small, medium, and large objects are defined as $(area<32^2)$, $(32^2<area<96^2)$ and $(area>96^2)$. Considering the area of most images is $640\times 480$, the areas of small, medium, and large objects in each image account for 0.33\%, 0.33\%$\sim$3\%, and 3\% of the entire image area\cite{COCO}. Notice that when the object vehicle is 17.98 meters away, it will be processed as a large object (i.e., 42000 pixels and account for 3\% of the entire image). While when the object is 163 meters away, it will be processed as a small object. Similar observation also applies to pedestrian detection. Considering the vision of cameras ranges from 20 meters to 200 meters\cite{Tesla, Audi}, the captured image will include objects with various areas.

Facing such complicated workloads, a single CNN model struggles to meet the requirements of accuracy and detection time when the ranges of sizes and areas of objects are wide. ~\textbf{~\textit{Tapping into heterogeneous CNNs is promising viability to mitigate this challenge.}} As shown in  Table~\ref{AP}, YOLO and SSD show various achieved Average Precisions (AP) for different object areas. We can observe that YOLO is good at small and medium object detection, while SSD is good at large object detection. Therefore, the object detection tasks in automated vehicles demand heterogeneous CNN models to ensure accuracy.

%\vspace{-0.9em} 
%\subsubsection{Heterogeneous Performance Requirements of Driving Automation Processing.}
%\vspace{-0.3em} 

\iffalse
\begin{table}[h!]
\renewcommand\arraystretch{1}
  \centering
  \scriptsize
  %\begin{tabular}{ |p{1cm}<{\centering}|p{1.8cm}<{\centering} |p{1.8cm}<{\centering} |p{1.8cm}<{\centering} |}
    \begin{tabular}{ p{2.1cm}<{\centering}p{0.5cm}<{\centering} p{0.5cm}<{\centering} p{0.5cm}<{\centering} p{0.5cm}<{\centering}p{0.5cm}<{\centering}p{0.5cm}<{\centering}}
    \hline
   \textbf{Function}&\textbf{FC}&\textbf{FLSC}&\textbf {RLSC}&\textbf{FRSC}&\textbf {RRSC}&\textbf {RC} \\
   \hline
    \textbf{Camera number} & 11&4&4&4&4&3\\
    \hline
    \textbf{Go straight(FPS)} & 40&30&20&30&20&10\\
    \hline
    \textbf{Turn left(FPS)}  & 40&40&30&30&20&10\\
    \hline
   \textbf{Reverse(FPS)} & 20&20&30&20&30&40\\
    \hline
  \end{tabular}
    \vspace{-1em} 
    \captionsetup{font={small}} 
    \caption{\textbf{Camera configurations in different scenarios. FC, FLSC, RLSC, FRSC, RRSC and RC represents forward, forward left side, rearward left side, forward right side, rearward right side adn rear cameras.} \label{ccids}}
\end{table}
\fi

\subsection{Variability of Performance Requirements in Driving Automation Tasks} 
Driving automation is a highly safety-critical application. Its safety guarantee relies on rich surrounding information collected by its intensively integrated cameras and sensors. 
%The study in \cite{CV} compares the number of cameras integrated in different car makers’ automated vehicles. 
Up until 2018, the number of cameras in Tesla is 8, while Audi, BMW and Mercedes Benz use 5 cameras\cite{CV}. Uber's self-driving Volvo SUV includes 7 cameras, and Ford Fusion includes 20 cameras to improve safety\cite{Uber}. In March 2020, Waymo unveiled its fifth-generation self-driving system\cite{Waymo}, which includes 29 cameras. According to this trend, it is reasonable to project that automated vehicles will deploy more than \textbf{30} cameras in the near future.

\begin{figure}[t] %h!
\centering
\includegraphics[width=2.4in,height=0.8in]{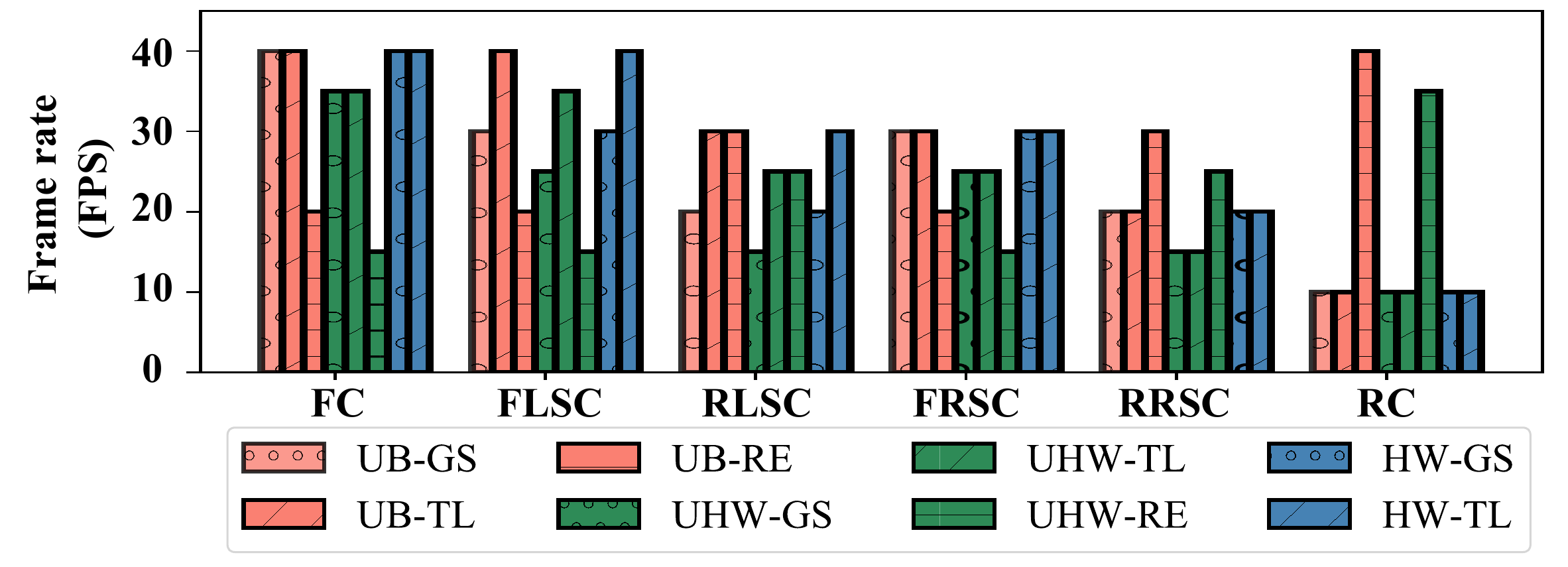}
\vspace{-1em} 
\captionsetup{font={small}}
\caption{\textbf{Frame rate requirements for different areas and scenarios.}\label{cfa}}
\end{figure}

However, these cameras can generate video streams at various image rates ranging from 10 to 40 frame-per-second (FPS) depending on its function \cite{yang2019re}, and this~\textbf{\textit{will incur differentiated performance requirements for the backend hardware accelerators}}. We will show this challenge using an example. As shown in Table~\ref{ccids}, we have 30 cameras which are configured into different function groups based on the configuration in Tesla, namely forward (FC), forward left side (FLSC), rearward left side (RLSC), forward right side (FRSC), rearward right side (RRSC) and rear cameras (RC). Figure~\ref{cfa} shows the required frame rates for each camera group under different driving areas, namely urban areas (UB), undivided-highway areas (UHW), and high-way areas (HW), and different driving scenarios, namely going straight (GS), turning left/right (TL), and reversing (RE). Notice that since reversing is not allowed on the highway, the corresponding frame rate of RC in HW is not provided.

%different functions to 30 cameras according to the distribution of the cameras in Tesla, and the frame rate of each camera in different scenarios are also assumed and given in Figure~\ref{cfa}. For instance, as for the forward camera, the frame rate of going straight and turning is the same in each area. And when a vehicle is reversing, the frame rate are all reduced in UB and UHW. However, because reversing is not allowed on the highway, the corresponding frame rate of FC in HW is not given here.

\begin{table}[t]
\renewcommand\arraystretch{1}
  \centering
  \scriptsize
  %\begin{tabular}{ |p{1cm}<{\centering}|p{1.8cm}<{\centering} |p{1.8cm}<{\centering} |p{1.8cm}<{\centering} |}
    \begin{tabular}{ p{2.1cm}<{\centering}p{0.5cm}<{\centering} p{0.5cm}<{\centering} p{0.5cm}<{\centering} p{0.5cm}<{\centering}p{0.5cm}<{\centering}p{0.5cm}<{\centering}}
    \hline
   \textbf{Function}&\textbf{FC}&\textbf{FLSC}&\textbf {RLSC}&\textbf{FRSC}&\textbf {RRSC}&\textbf {RC} \\
   \hline
    \textbf{Camera number} & 11&4&4&4&4&3\\
    \hline
  \end{tabular}
    \vspace{-1em} 
    \captionsetup{font={small}} 
    \caption{\textbf{Configuration of cameras with different functions.} \label{ccids}}
\end{table}

\begin{table}[t]
\renewcommand\arraystretch{1}
 %\vspace{-1.5em} 
  \centering
  \scriptsize
  %\begin{tabular}{ |p{1cm}<{\centering}|p{1.8cm}<{\centering} |p{1.8cm}<{\centering} |p{1.8cm}<{\centering} |}
    \begin{tabular}{ p{1.8cm}<{\centering}p{0.6cm}<{\centering} p{0.6cm}<{\centering} p{0.6cm}<{\centering}p{0.6cm}<{\centering}p{1cm}<{\centering}  }
    \hline
   &\textbf{DET}&\textbf{TRA}&\textbf {YOLO}&\textbf {SSD}&\textbf {GOTURN} \\
    \hline
    \textbf{Go straight(FPS)} & 870&840&435&435&840\\
    \hline
    \textbf{Turn left(FPS)}  & 950&920&475&475&920\\
    \hline
   \textbf{Reverse(FPS)} &740&740&370&370&740\\
    \hline
  \end{tabular}
    \vspace{-1em} 
    \captionsetup{font={small}} 
    \caption{\textbf{The performance requirements of vehicles in urban area. The requirement of turning right is same as turning left.} \label{tpr}}
\end{table}

Then, we can infer the requirements of image processing capability for each perception task in different areas. For example, Table~\ref{tpr} shows the FPS requirements for YOLO, SSD, and GOTURN in urban area under three driving scenarios. In going straight scenario, since object detection (DET) should be executed for all images from the cameras, the performance requirement is 870 FPS for the backend accelerator. However, since the object tracking is not supposed to perform for the rear cameras, the total performance requirement for object tracking (TRA) is 840 FPS. For the DET task, we use YOLO for small and medium object detection and use SSD for large object detection, hence the average processing capability for YOLO and SSD is 435 FPS. We use GOTURN to process the TRA task and the processing capability is 840 FPS. 

%As YOLO is good at small and medium object detection, while SSD is good at large object detection, here we assume that for an image from each camera, YOLO and SSD will be used alternately for processing object detection tasks. At the same time, we use GOTURN to process the object tracking task. Therefore, for the scenario of going straight, the hardware platform in the autonomous driving vehicle needs to process 435 frames YOLO, 435 frames SSD and 840 frames GOTURN per second.  

We can observe that when the driving area or driving scenario changes, the FPS requirement for each camera in a vehicle will change consequently. Therefore, the amount of mixed CNN models that a vehicle needs to process within a certain period also will change accordingly. ~\textbf{\textit{Such variability in performance requirements and workload mix should be carefully handled by computing platform}}.
\section{Design Challenges of Driving Automation System}
Both industry\cite{Audi,Tesla,Waymo} and academia\cite{UCSan,Illinois,Texas,Safety} are intensively exploring the driving automation system.The Society of Automobile Engineers (SAE)\cite{driving2014levels} defines six levels of driving automation from Level 0 to Level 5, where Level 5 is fully autonomous. To deliver a functional and practical driving automation system with a safety and reliability guarantee, designers should carefully analyze and optimize every technical detail from hardware through software. In this section, we explore the system design challenges imposed by the variability of workloads and performance requirements in driving automation. These challenges motivate us to propose a comprehensive framework that synergistically handles the heterogeneous hardware accelerator design principles, system design criteria, and task scheduling mechanism. 

\subsection{Hardware Platform Design}
\noindent\textbf{Multi-accelerator is Needed for Driving Automation Processing.} The cameras and sensors on automated vehicles can generate a massive amount of data for real-time analysis. We show the relationship between the speed of the car and the requirements of the frame rates in Table~\ref{frameRate}, which are collected from multiple studies. We can observe that the required frame rate is around 20 FPS (frames per second). Note that a high frame rate is necessary for high-speed driving\cite{2019survey}. In KITTI the max frame rate for a speed of 90km/h could be 100 FPS. In industry practice, Audi sets the camera frame rate in driver assistance systems as 25 FPS \cite{audifps}. The Tesla Model 3 adopts 36 FPS\cite{Teslafps}. With the much higher safety requirements in the fully driving automation, the camera frame rates will be greater than \textbf{40} FPS in the future.

\begin{table}[t]
\renewcommand\arraystretch{1}
 %\vspace{-1.5em} 
  \centering
  \scriptsize
  \begin{tabular}{ p{3cm}<{\centering}p{2.1cm}<{\centering} p{1.9cm}<{\centering}}
  \hline
    &\textbf{Max velocity(km/h)} &\textbf{Frame rate(FPS)} \\
    \hline
    \textbf{KITTI}\cite{KITTI2}&90&10-100\\ \hline
    \textbf{ApolloScape}\cite{apolloscape}&30&30\\ \hline
   \textbf{Princeton}\cite{Princeton}&80&10\\ \hline
   \textbf{VisLab}\cite{VisLab}&70.9&>25\\ \hline
   \textbf{Oxford RobotCar}\cite{Oxford}&Not Mentioned&11.1-16\\ \hline
   \textbf{Comma.ai}\cite{comma}&Not Mentioned&20 \\
    \hline
  \end{tabular}
    \vspace{-1em} 
    \captionsetup{font={small}} 
    \caption{\textbf{The camera frame rates in different researches.}\label{frameRate}}
\end{table}

\begin{table}[t]
\renewcommand\arraystretch{1}
  \centering
  \scriptsize
  \begin{tabular}{ p{2.6cm}<{\centering}p{2.4cm}<{\centering} p{2cm}<{\centering}}
  \hline
    \textbf{Device type}&\textbf{YOLO type} &\textbf{Frame rate(FPS)} \\
    \hline
    \textbf{GTX TitanX}\cite{yolo9000}&Sim-YOLO-v2&88\\ \hline
    \textbf{GTX TitanX}\cite{redmon2016you}&FAST YOLO&155\\ \hline
    \textbf{Zynq UltraScale+}\cite{preusser2018inference}&Tincy YOLO&30\\ \hline
    \textbf{Zynq UltraScale+}\cite{2018lightweight}&Lightweight YOLO-v2&40.81\\ \hline
   \textbf{Virtex-7 VC707}\cite{nguyen2019high}&Tiny YOLO-v2&66.56\\ \hline
   \textbf{Virtex-7 VC707}\cite{nguyen2019high}&Sim-YOLO-v2&109.3\\ \hline
   \boldsymbol{$ADM-7V3 FPGA_1$}\cite{ding2019req}&Tiny YOLO&208.2\\ \hline
   \boldsymbol{$ADM-7V3 FPGA_2$}\cite{ding2019req}&Tiny YOLO&314.2\\ \hline
  \end{tabular}
    \vspace{-1em} 
    \captionsetup{font={small}} 
    \caption{\textbf{Peak frame rates when run ML models on single accelerator.}\label{CYOLO}}
\end{table}

\begin{table}[!htpb]
\renewcommand\arraystretch{1}
%\vspace{-1em} 
  \centering
  \scriptsize
  %\begin{tabular}{ |p{1cm}<{\centering}|p{1.8cm}<{\centering} |p{1.8cm}<{\centering} |p{1.8cm}<{\centering} |}
    \begin{tabular}{ p{0.7cm}<{\centering}p{1.9cm}<{\centering} p{2cm}<{\centering} p{2cm}<{\centering} }
    \hline
   &\textbf{SconvOD (FPS)}&\textbf{SconvIC (FPS)}&\textbf {MconvMC (FPS)} \\
    \hline
    \textbf{YOLO} & 170.37&132.54&149.32\\
    \hline
    \textbf{SSD}  & 74.99&82.94&82.57\\
    \hline
   \textbf{GOTURN} & 352.69&350.34&500.54\\
    \hline
  \end{tabular}
    \vspace{-1em} 
    \captionsetup{font={small}} 
    \caption{\textbf{The performance of various accelerators.} \label{various_time}}
\end{table}

\begin{table}[!htpb]
\renewcommand\arraystretch{1}
  \centering
  \scriptsize
  %\begin{tabular}{ |p{1cm}<{\centering}|p{1.8cm}<{\centering} |p{1.8cm}<{\centering} |p{1.8cm}<{\centering} |}
    \begin{tabular}{ p{1.4cm}<{\centering}p{1.5cm}<{\centering} p{2.2cm}<{\centering} p{1.5cm}<{\centering}  }
    \hline
   &\textbf{YOLO}&\textbf{SSD}&\textbf {GOTURN} \\
    \hline
    \textbf{Go straight} & (1 SO, 2 SI)&(3 SO, 1 SI, 2 MM)&(1 SI, 1 MM)\\
    \hline
    \textbf{Turn left}  &(2 SO, 1 MM)&(2 SO, 4 SI)&(2 MM) \\
    \hline
   \textbf{Reverse} &(3 SI)&(2 SO, 3 MM)&(2 SO, 1 SI)\\
    \hline
  \end{tabular}
    \vspace{-1em} 
    \captionsetup{font={small}} 
    \caption{\textbf{The task allocation in (4 SconvOD, 4 SconvIC, 3 MconvMC). SO, SI and MM means SconvOD, SconvIC and MconvMC respectively.} \label{443}}
\end{table}

With multiple cameras and high frame rate generation for each camera, a single automated car can generate enormous data and overwhelm the processing capability of a single accelerator. We quantify the capability of emerging accelerators in terms of peak frame rates, as shown in Table~\ref{CYOLO}. We run the state-of-the-art object detection application YOLO and its derivations\cite{yolo9000} on single accelerators. Note that though for some accelerators, the peak FPS can reach 314.2, it still cannot meet the maximum processing requirements of 1200 FPS. This is calculated based on the assumption of a car with 30 cameras and a generation rate of 40 FPS for each camera. Considering that Tesla FSD\cite{FSD} has integrated four accelerators that can process 2300 frames per second, we can expect that future computing platforms for the automation system will base on multi-accelerator architecture.

\noindent\textbf{Heterogeneous hardware is Needed for Driving Automation Processing.}
Here, we design three representative CNN accelerators (SconvOD, SconvIC, and MconvMC) based on our CNN accelerator taxonomy (detail in Section 5.1). By using these CNN accelerators, we can construct different homogeneous and heterogeneous platforms to process dynamic performance requirements in an autonomous driving system.

\begin{figure}[t] %h!
\centering
\includegraphics[width=3.2in,height=1.8in]{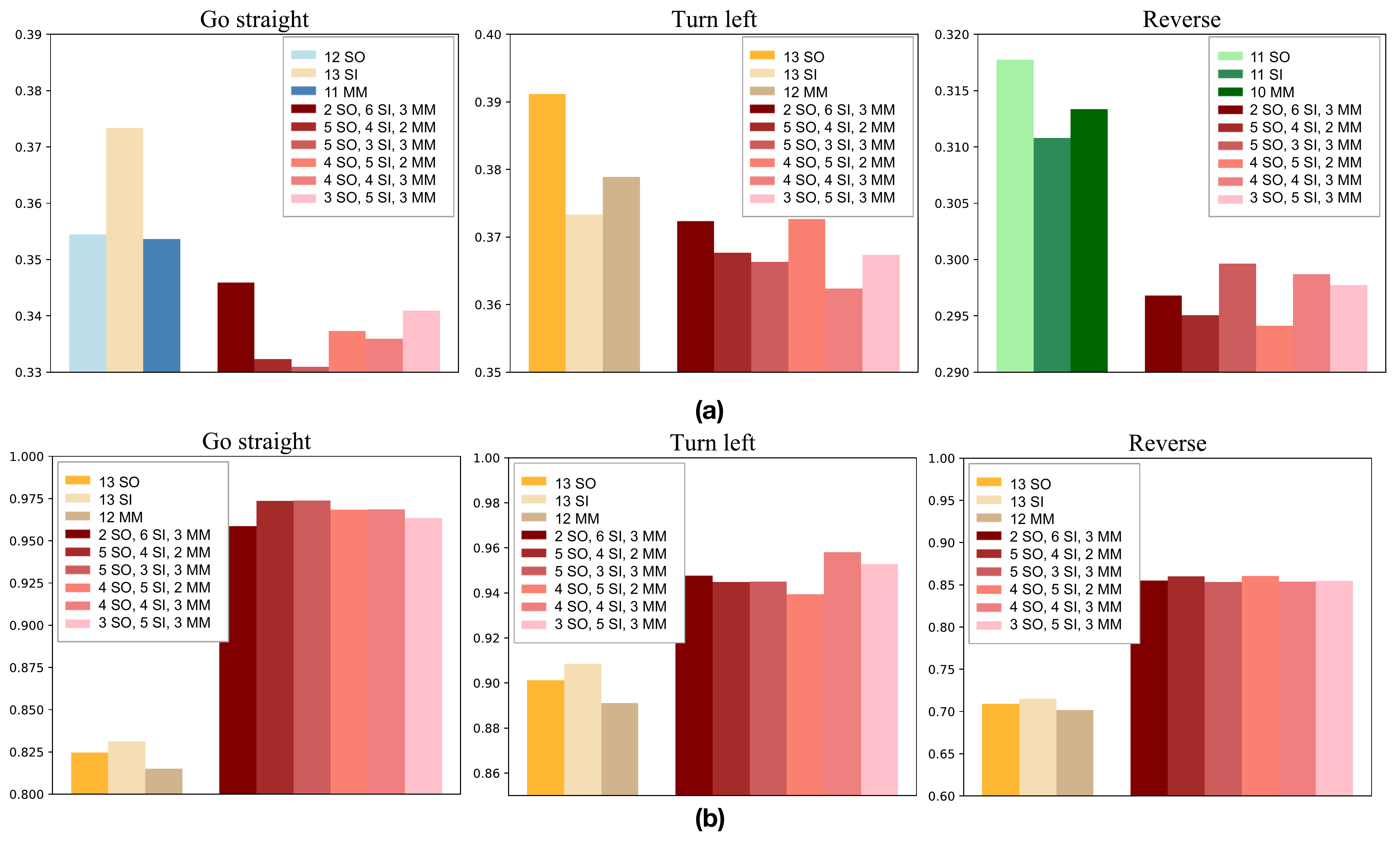}
\vspace{-1em} 
\captionsetup{font={small}}
\caption{\textbf{Comparison of (a) energy consumption and (b) resource utilization rate between homogeneous and heterogeneous platforms. SO, SI and MM means SconvOD, SconvIC and MconvMC, respectively.}\label{newhohe2}}
\end{figure}

For homogeneous platforms, according to the performance given in Table~\ref{various_time}, we can obtain the number of accelerators required by each homogeneous platforms in different environments. Now, we will discuss the best configuration of each homogeneous platform under an assumption that a vehicle only travels in urban areas. When a vehicle is going straight, in order to meet the  performance requirements in Table~\ref{tpr}, for a homogeneous platform based on SconvOD, 3 SconvOD, 6 SconvOD, and 3 SconvOD are needed to process YOLO, SSD and GOTURN respectively, thus this homogeneous platform must contain 12 SconvOD. In the legend of Figure~\ref{newhohe2} (a), the most suitable accelerator number of each homogeneous platform in different scenarios are given. Figure~\ref{newhohe2} (a) shows the energy consumption of each homogeneous platform. 

Since the hardware platform in the vehicle is fixed in advance, it needs to meet the performance requirements of all scenarios. For the homogeneous platforms based on the same accelerator in Figure~\ref{newhohe2} (a), we will choose the maximum number of accelerators in all scenarios to construct the final homogeneous platform. For instance, if the homogeneous platform is composed of SconvOD, the final platform will include 13 SconvOD. In Figure~\ref{newhohe2} (b), we show the resource utilization rate of different final homogeneous platforms in all scenarios.

When the heterogeneous platform (4 SconvOD, 4 SconvIC, 3 MconvMC) with the task allocation in Table~\ref{443}, in Figure~\ref{newhohe2} (a), we can find that the energy of all heterogeneous platforms is lower than that of homogeneous platforms in the same scenarios. In Figure~\ref{newhohe2} (b), the resource utilization of this heterogeneous platform is 96.86\%, 95.81\% and 85.40\% for go straight, turn left and reverse, respectively, which are also higher than all homogeneous platforms. Therefore, in autonomous driving systems, heterogeneous platforms which can not only
keep the lower energy consumption, but also achieve the highest resource utilization in all scenarios. It should be noted that, there are many methods to schedule tasks on the same heterogeneous platform. In Figure~\ref{newhohe2}, we use the best method on each heterogeneous platform. The best method can bring the maximum geometric mean of resource utilization rate among three scenarios for each heterogeneous platform.

Moreover, considering the driving automation system is still evolving, a heterogeneous accelerator architecture can better accommodate the ever-changing new algorithms and applications in this area\cite{wang2017}.

\subsection{System Design Criteria} To efficiently process such a large amount of CNN-based tasks with high variability on the complicated hardware substrate, effective criteria for system design that are tailored to driving automation should be defined. Obviously the overall performance of the computing platform should be considered at first. Specifically, the execution time, energy consumption, and resource utilization of the platform are expected to be optimal after all tasks have been processed. 

%two major issues should be taken into consideration when these tasks are processed on a multi-accelerator hardware platform. Firstly, the overall performance of the hardware platform. After all the tasks have been processed, three metrics, running time, energy consumption, and resource utilization of the platform, are expected to be optimal. 

Another indispensable criterion is the safety requirement for the task processing. The computing platform needs to provide differentiated processing time for the object recognition or object tracking tasks from different cameras. 
%For different cameras, due to a variety of maximum distances, the hardware platform needs to provide different processing time when performing object recognition or object tracking on pictures from different cameras. 
For instance, the detection task of an object in front of a vehicle with a distance of 50 meters has higher priority than the object that is 80 meters away. Therefore, we need to find a metric to describe whether the hardware platform’s processing time for tasks from each camera is safe. A detailed discussion of system design criteria is provided in Section 6.

\subsection{Scheduling Mechanisms}
As we know, task mapping on hardware substrate is an NP-complete problem and is generally solved using heuristic\cite{11,ATA,H1,H2,H3,H4,H5,EDP} or guided random-search-based algorithms\cite{hou,shroff,correa,Haluk}. However, the scheduling strategies based on these algorithms fail to see the global situation of computing platform such as current resource utilization, the longest execution time among all cores, which often results in a suboptimal allocation. An efficient task scheduling mechanism is the crux to trade-off the metrics that are defined in the system design criteria. We will elaborate our choice in Section 7. 

%how to designate tasks to different accelerators in HMAI needs to be carefully designed.
%As we know, the max distance of each camera on a vehicle is different. For example, the max distance of Tesla's narrow forward camera, forward side camera and rear camera are 250m, 80m and 50m respectively. Therefore, when there are objects 250 meters in front of the car, 80 meters away from the side, and 50 meters to the rear, in order to prevent active and passive collisions, the images generated by each camera at this moment need to be processed within a suitable time range respectively. That is, for objects in each direction, autonomous vehicles should make reasonable actions in time. According to the RSS model\cite{Save}, we can get the maximum value of the suitable time range of each camera (detail will be given in Section 5.1), and hardware platform’s processing time for images from each camera cannot be greater than the corresponding maximum value for security. Therefore, we need to find a metric to describe whether the hardware platform’s processing time for tasks from each camera is safe.

%According to the four metrics from the two issues aforementioned, an autonomous driving system can guide the task execution on the hardware platform. Furthermore, the scheduling mechanism should be applied in the guideline.

\section{A Synergistic Framework}
\begin{figure}[t]  
\centering
\includegraphics[width=2.8in,height=1.8in]{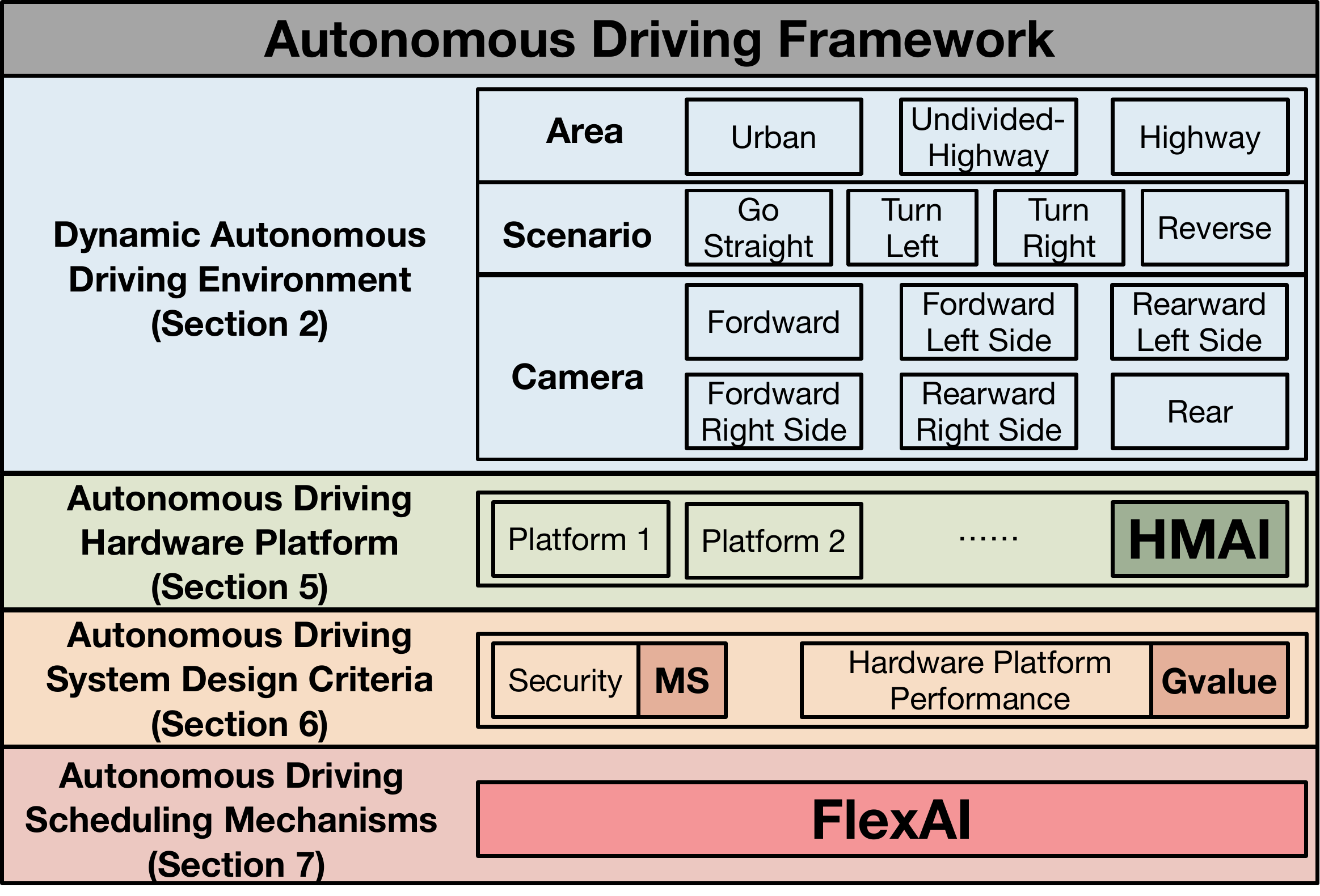}
\vspace{-1em} 
\captionsetup{font={small}} 
\caption{\textbf{An overview of Framework in Autonomous Driving System.} \label{Framework}}
\end{figure}
We propose a synergistic framework for driving automation to bridge the gap between variable driving automation workloads and complicated hardware substrates, as shown in Figure~\ref{Framework}. Specifically, we first propose a CNN taxonomy and design principles for hardware accelerators. Based on these knowledge, we propose a novel heterogeneous multi-core AI accelerator (HMAI) to provide the hardware substrate for the driving automation tasks with variability. Our framework also defines the system design criteria to better utilize hardware resources and achieve increased throughput while satisfying the performance and energy restrictions. Specifically, we propose two metrics, Matching Score (MS) and Global State Value (Gvalue) to formalize the criteria. MS pays attention to the safety requirements of various tasks in driving automation systems, while Gvalue puts more weight onto the overall performance of HMAI that reflects the globality. Finally, our framework employs a deep reinforcement learning (RL)-based task scheduling mechanism FlexAI to meet the system design criteria. 

%the dynamic performance requirements from autonomous driving environment. Then in order to process a large number of continuous and dynamic CNN-based tasks in the driving automation system, we design a heterogeneous multicore AI platform HMAI for this framework. Next, by designing two metrics in system design criteria: Matching Score (MS) paying attention to safety, and Global State Value (Gvalue) reflecting hardware platform performance, autonomous driving system can guide the task execution on the platform. Finally, by comparing with multiple algorithms, we choose to use reinforcement learning in our task scheduling mechanisms (FlexAI).

\section{HMAI-A Heterogeneous Multicore AI Platform}
In this section, we propose a heterogeneous multicore AI platforms (HMAI) tailored to the CNN-related perception tasks for driving automation framework. To achieve this goal, we first propose a taxonomy of existing accelerator architectures for CNNs and choose the best sub-accelerators for driving automation workloads based on the taxonomy. We then propose the architecture of HMAI that consists of three representative sub-accelerators.

\subsection{A Taxonomy for CNN Accelerators}

\begin{figure*}[ht] %h!
\centering
\vspace{-1.5em} 
\includegraphics[width=6.6in,height=1in]{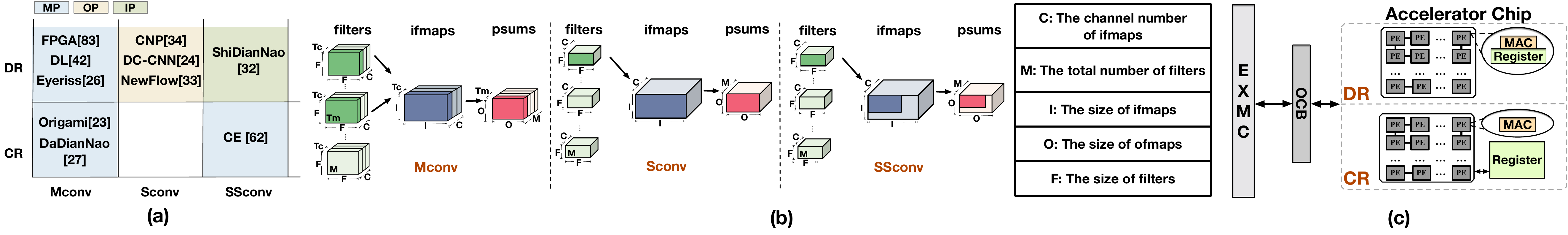}
\vspace{-1em} 
\captionsetup{font={small}}
\caption{\textbf{(a) Quadrant Classification for CNN accelerators\cite{ShiDianNao, CNP, NeuFlow,Dyna,CE,Eyeriss2,FPGA,DL,Origami,Dadiannao}. (b) Data processing style. (c) Register allocation.}\label{Classification}}
\vspace{-1em} 
\end{figure*}

\begin{figure*}[h]%htbp
\centering
\begin{minipage}[t]{0.48\textwidth}
\centering
\includegraphics[width=3.3in,height=1.15in]{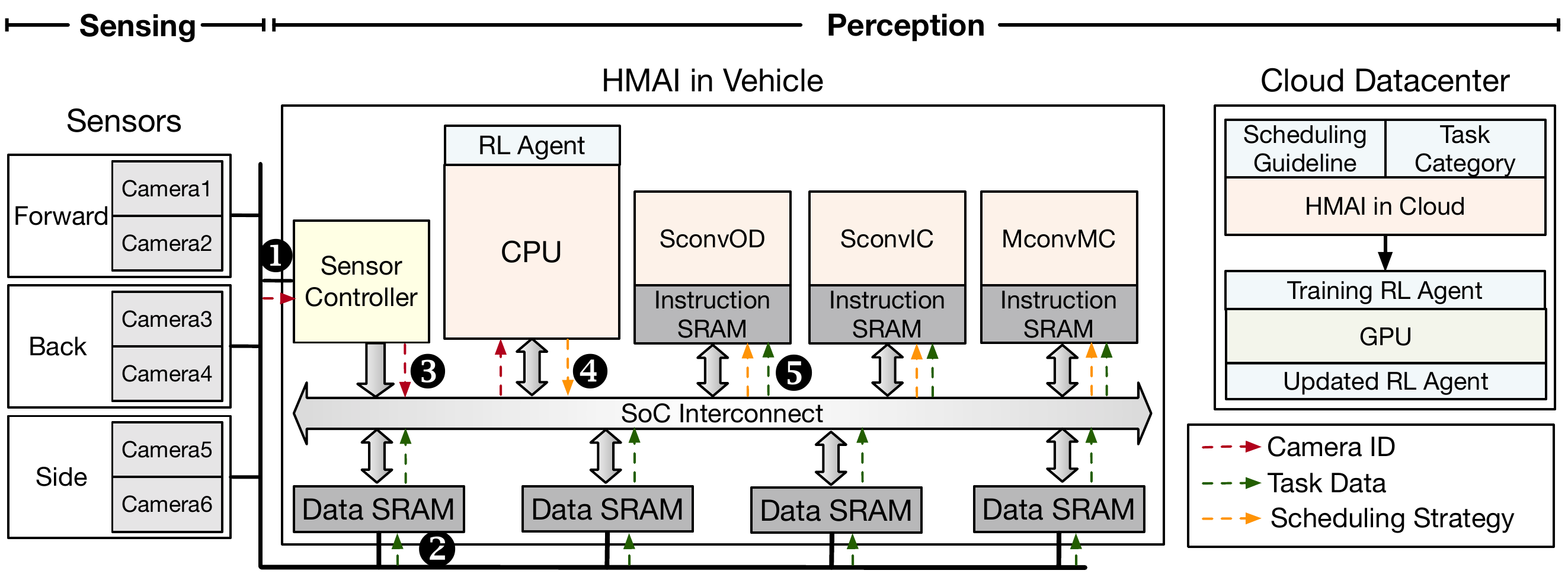}
\vspace{-1em} 
\captionsetup{font={small}}
\caption{\textbf{An overview of HMAI.} \label{chip}}
\end{minipage}
\begin{minipage}[t]{0.48\textwidth}
\centering
\includegraphics[width=3.3in,height=1.15in]{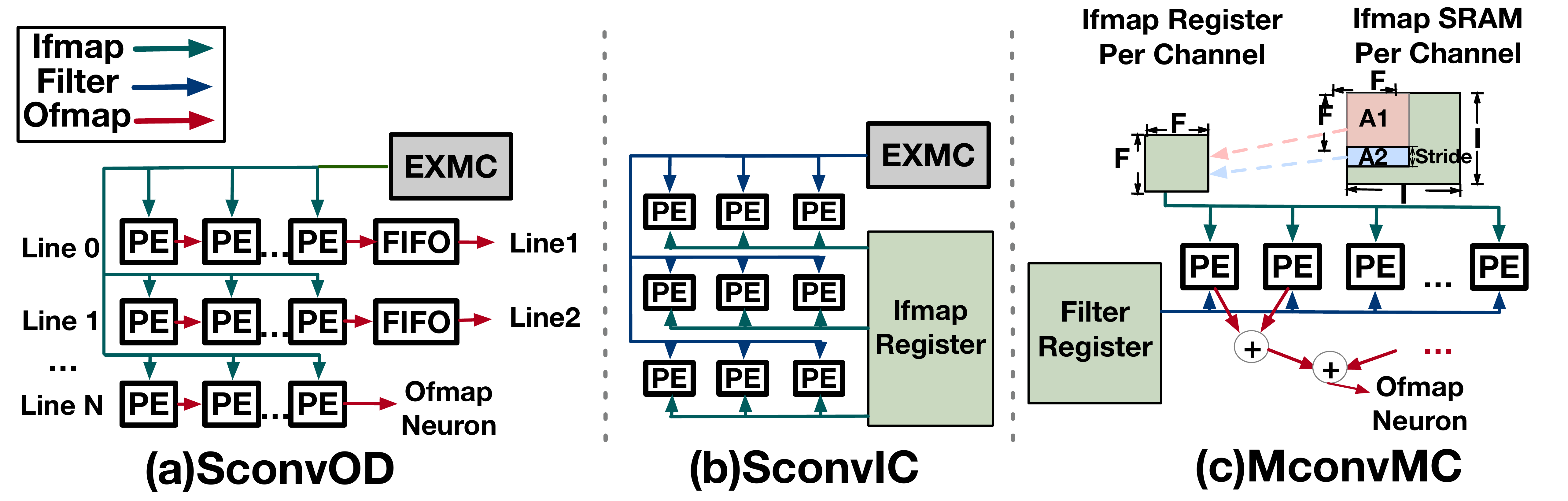}
\vspace{-1em} 
\captionsetup{font={small}}
\caption{\textbf{SconvOD, SconvIC and MconvMC in HMAI.} \label{3ac}}
\end{minipage}
\vspace{-1.5em}
\end{figure*}

To choose the representative sub-accelerator architectures for HMAI, we need a comprehensive understanding of existing CNN accelerators as shown in Figure~\ref{Classification} (a). We propose a taxonomy for emerging CNN accelerators with respect to data processing style, register allocation, and data propagation types.

\iffalse
A CNN network mainly consists of three layers: convolutional layers (CONV), pooling layers (POOL), and fully-connected layers (FC). These layers contain three data types, input feature map (ifmaps), filters, and output feature map (ofmaps), which are generated by accumulating the partial sums (psums). A detailed description for these data types is given in Table~\ref{ParaCNN}.
\begin{table}[h]
\renewcommand\arraystretch{1}
    \vspace{-0.8em} 
  \centering
  \scriptsize
  \begin{tabular}{ p{1cm}<{\centering}p{3cm}<{\centering} }
    \hline 
    \textbf{Parameter} & \textbf{Description} \\
    \hline
    \textbf{C} & The channel number of ifmaps\\
    \hline
    \textbf{M} & The total number of filters\\
    \hline
   \textbf{I} & The length or width of ifmaps\\
    \hline
    \textbf{O} & The length or width of ofmaps\\
    \hline
    \textbf{F} & The length or width of filters\\
    \hline
  \end{tabular}
    \vspace{-1em} 
    \captionsetup{font={small}}
    \caption{\textbf{The parameters of CNN.} \label{ParaCNN}}
\vspace{-1em}
\end{table}
\fi

\noindent\textbf{Data Processing Style.} In this work, we first categorize the CNN accelerators into three styles as shown in Figure~\ref{Classification} (a), namely S(ingle)conv, S(pecial)Sconv and M(ultiple)conv according to their data processing methods. As shown in Figure~\ref{Classification} (b), Sconv processes a whole 2D convolution each iteration. While SSconv only processes a part of 2D convolution each iteration.
For Mconv, it processes multiple 2D convolutions each iteration. We define the data processed by the accelerator in each iteration as a basic calculation unit (a.k.a., \textbf{BasicUnit}). For example, in Figure~\ref{Classification} (b), the size of filters in a BasicUnit of Mconv is $F\times F\times T_{m}\times T_{c}$, the size of ifmaps is $I\times I\times T_{c}$, and the size of psums is $O\times O\times T_{m}$.

%And in Mconv, the computation of one CNN layer is divided into multiple blocks, where each block contains $T_{c}$-channel ifmaps and $T_{m}$ filters can be processed each time.

%{ShiDianNao\cite{ShiDianNao}, CNP\cite{CNP}, NewFlow\cite{NeuFlow}, DC-CNN\cite{Dyna}, CE\cite{CE},Eyeriss\cite{Eyeriss2}, FPGA\cite{FPGA}, DL\cite{DL}, Origami\cite{Origami}} Dadiannao\cite{Dadiannao}

\noindent\textbf{Register Allocation.} Figure~\ref{Classification} (c) illustrates a high-level block diagram of a typical CNN accelerator. It consists of an accelerator chip and an external memory chip (EXMC). Processing elements (PE) array is often used as the main functional component in the accelerator chip, which contains multiply-accumulate unit (MAC) as computation units. The on-chip buffer (OCB) is used to store ifmaps, filters, and psums. We classify the CNN accelerators into two categories with respect to register type: dispersive register (DR) and concentrated register (CR). In DR the registers are dispersed in each PE, while in CR a centralized storage is used and never stores psums. The size requirements of these registers are different among accelerators. Table~\ref{DiffStru} lists different structure designs for Sconv, SSconv and Mconv.

\begin{table}[t]
\renewcommand\arraystretch{1}
\scriptsize
\centering
\begin{tabular}{p{1.7cm}<{\centering}p{0.6cm}<{\centering}p{0.4cm}<{\centering}p{1.4cm}<{\centering}p{2.4cm}<{\centering}}
\hline
\textbf{} & \textbf{EXMC} & \textbf{OCB} &  \textbf{PE Register}& \textbf{MAC num in each PE} \\ \hline
\textbf{Sconv \& SSconv} &\checkmark &$\times$ &CR or DR & 1 	\\ \hline
\textbf{Mconv} &\checkmark&\checkmark &CR or DR & >1  \\ \hline
\end{tabular}
  \vspace{-1em} 
 \captionsetup{font={small}}
\caption{\textbf{Different structure design for Sconv, SSconv and Mconv.} \label{DiffStru}}
\end{table}

\noindent\textbf{Data Propagation Types.} We further define three data propagation types for the data propagation between different PEs, Ofmaps Propagation (OP), Ifmaps Propagation (IP), and Multiple Propagation (MP). (1) For OP, the ifmaps are directly sent to PEs in one BasicUnit. The filters are fixed in the PEs in advance. The psums are accumulated during the data propagation between PEs. Upon all PEs are traversed, the ofmaps neuron is generated. (2) For IP, in one BasicUnit, the filters will be sent directly to the PEs. The ofmaps are fixed in PEs. The ifmaps propagate between PEs for reuse. (3) For MP, in one BasicUnit, there will be single or multiple types of data propagation between PEs. Figure~\ref{Classification} (a) shows typical CNN accelerators with different data propagation types. Note that data propagation always indicates data transfer between PEs' registers. If there is no register in each PE, data propagation means data transfer between PE array and CR, since CR is equivalent to a collection of registers in each PE.
%We can observe that the data propagation type of Mconv generally is MP, while Sconv with DR usually is OP, and IP and MP both exist in SSconv. 

%(2) When filters are propagated between PE, another type of data will be propagated as well because the size of filters is relatively small. 

%\vspace{-1em} 

%Meanwhile, in most CNNs \cite{AlexNet,VGG,GoogleNet,LSTM}, the calculation amount and the input/output data amount of CNN layers and FC layers account for a large proportion of entire forwarding operations. Hence, we are currently modeling CONV layers in CNN (FC layer can be replaced by a 1$\times$1 CNN layer), and the mechanism proposed here can be extended to estimate layers of any kind.
\vspace{-0.4em} 
\subsection{The Architecture of HMAI}
Based on the above mentioned CNN accelerator taxonomy, we propose a heterogeneous multicore AI accelerator (HMAI) that contains three representative accelerators (SconvOD, SconvIC, MconvMC) in an automated vehicle. Each AI core has its specific architecture, as shown in Figure~\ref{3ac}. Figure~\ref{chip} shows the architecture and data/control flows in the HMAI: \ding{182}cameras with different frame rate in the sensing component will generate multiple frames each time. After the cameras generate the frames data, they will signal the sensor controller with their ID separately and the sensor controller will launch DMA transition between camera and Data SRAM; \ding{183}since each camera has its identical data SRAM, DMA will transfer the frame point to point, from a specific camera to its related data SRAM;
\ding{184}at the same time, CPU will acquire camera ID of current task from the controller through SoC interconnect; \ding{185} the well-trained RL agent in the CPU will generate a scheduling strategy for all tasks according to its camera ID, and then this strategy will be sent to a corresponding accelerator through interconnect (the RL agent can be retrained by GPU in cloud data center, when the task category and scheduling strategy need to be changed); \ding{186}guided by the scheduling strategy, each accelerator will read the frame data from the corresponding SRAM and start the computation.

\noindent\textbf{Why these accelerators?} The HMAI is tailored to the CNN-related driving automation perception tasks. We choose to implement all data processing styles, the Sconv, SSconv and Mconv as described in Section 5.1. To cover all data propagation types in HMAI, we further choose to implement Sconv-OP, SSconv-IP and Mconv-MP based on multiple existing accelerator types as shown in Figure~\ref{Classification}. To cover the register allocation methods, we implement SconvOD as Sconv-OP-DR, SconvIC as SSconv-IP-CR and MconvMC as Mconv-MP-CR. 

\noindent\textbf{The architecture design of sub-accelerators.} As shown in Figure~\ref{3ac} (a), SconvOD is based on NewFlow~\cite{NeuFlow}. In SconvOD, each ifmaps neuron only needs to be taken from the EXMC once. In each cycle, the same ifmaps neuron is sent to all PEs, but not every PE will generate a valid signal for this ifmaps. Different filter weights are fixed in different PE's registers in advance. As to ofmaps neurons, it will be obtained after propagating to all PEs and FIFOs. The design of SonvIC is shown in Figure~\ref{3ac} (b), which is based on ShiDianNao~\cite{ShiDianNao}. In each cycle, the same filter weight is sent to all PEs. Different ifmaps neurons are read from the ifmaps register (the ifmaps register has the double buffer) to different PEs. Each PE computes only one output neuron each time. 
%%In AC2, every 9 output neurons are calculated, all weights of a filter need to be re-read, so when the size of the filter is greater than 1, the performance of AC2 will be greatly affected. Because the filter size of all layers in GOTURN is greater than 1, the performance of AC2 processing GOTURN is significantly lower than the other two accelerators.
The design of MconvMC is based on Origami~\cite{Origami}, and it is shown in Figure~\ref{3ac} (c), where its parameters of BasicUnit $T_{m}= T_{c}$. In ifmaps SRAM A1, the neurons of $T_{c}$ channels are sent to ifmaps register. Then the register sends ($F\times F\times T_{c}$)-size data to the PE array. Each PE will receive ($F\times F$)-size data, while the data in A2 will be sent to the register. For filters, different $F\times F$ are sent to different PEs at each cycle until all corresponding $T_{m}$-size filters are sent. In order to guarantee the pipelining, each PE will produce a result of matrix multiplication, and then the results in all PEs will be accumulated and sent out.

%%%%%%%%%%%%%%%%%%%important DO NOT DELETE%%%%%%%%%%%%%%%%%%%%%%%%%%
%The the frame rate of AC1, AC2 and AC3 for processing YOLO is 144FPS, 132FPS and 143FPS respectively. \cite{TAIAD} discover it is critical for the autonomous driving systems to be able
%to finish perception tasks at a frame rate higher than 10FPS, and the mean frame rate in ASIC to process perception tasks is 11FPS. Besides, in light of the most YOLO accelerator performance in Table~\ref{CYOLO}, our accelerators performance are acceptable.
%%%%%%%%%%%%%%%%%%%%%%%%%%%%%%%%%%%%%%%%%%%%%%%%%%%%%%%%%%%%%%%%%%%%%%%%%%%

%\begin{figure}[h]
%\centering
%\vspace{-1em}
%\includegraphics[width=2.8in,height=0.8in]{ISCA2019-latex-template/figure/ACs.pdf}
% \vspace{-1em} 
% \captionsetup{font={small}}
%\caption{\textbf{(a) ACs cycle. (b) ACs energy.} %\label{acs}}
% \vspace{-1em} 
%\end{figure}
% \vspace{-1em} 

\vspace{-0.5em} 
\section{System Design Criteria}
In this section, we propose Matching Score (MS) and Global State Value (Gvalue), to assist autonomous driving system guide the task execution on platforms.

\subsection{Matching Score (MS)}
Like Tesla (with 8 cameras), automated vehicles normally equip with multiple surrounding cameras to receive 360 degrees of visibility. As the different cameras have different max distance\cite{Tesla}, each camera has its own requirement for their response time. This \textbf{response time} means automated vehicle's processing time for each camera's task. Based on the different camera's response time, we define their matching score (MS). Specifically, we characterize the camera's MS under object detection and object tracking tasks.
%the time to process object detection or object tracking for the image from each camera by the automated vehicle. According to the different camera's response time, we define their matching score (MS). In this section, we first characterize the MS when the automated vehicle processes object detection for the image from each camera, then the MS for object tracking is introduced.

\noindent\textbf{Matching Score - Object Detection}  
Cameras in vehicles can be divided into three categories: forward, rear and side cameras. We first introduce the MS of forward cameras. When an obstacle is detected by a forward camera, this obstacle may be in one of the three states: (1) moving in the same direction as the vehicle, (2) standing still, and (3) moving in the opposite direction as the vehicle. Among those, the obstacle in the third state needs the shortest time to be detected, and we define this shortest time as the \textbf{safety time} of this forward camera.

%Therefore, when a vehicle processes object detection for images from one of the forward cameras, the response time must be less than relevant camera's safety time. 

%%%%%%%%%%%%%%%%%%%%%%%%%%%%%%%%important--do not delete%%%%%%%%%%%%%%%%%%%%%

Based on the Responsibility-Sensitive Safety (RSS) safety model\cite{Save}, the safety time of each camera can be derived. RSS reveals the relationship between safe distances and processing time of vehicles in different scenarios. When the two vehicles are driving at opposite directions, \cite{Save} proposes the minimal safe distance $d_{min}$ between rear car $c_{1}$ and front car $c_{2}$ with velocities $v_{1}$ , $v_{2}$.

\vspace{-1em} 
\begin{equation}
\scriptsize
d_{min}=\left[ \frac{v_{1}+v_{1,\rho}}{2}\rho + \frac{v_{1,\rho}^{2}}{2a_{min,brake,correct}} +  \frac{\lvert v_{2}\rvert+v_{2,\rho}}{2}\rho + \frac{{v_{2,\rho}}^{2}}{2a_{min,brake}} \right]\label{safedis}
\end{equation}
\vspace{-1em} 

In Equation(\ref{safedis}), $\rho$ is the processing time of $c_{1}$, $a_{max,accel}$ is vehicle's acceleration, $v_{1,\rho}=v_{1}+\rho a_{max,accel}$, and $v_{2,\rho}=\lvert v_{2}\rvert +\rho a_{max,accel}$. $a_{min,brake,correct}$ and $a_{min,brake}$ are the breaking acceleration of $c_{1}$ and $c_{2}$ respectively.

In this paper, we set $d_{min}$ to the max distance of each camera. $v_{1}$ and $v_{2}$ are the maximum velocity allowed in different areas (the maximum velocity in urban areas is 60km/h, undivided-highways is 80km/h and highways is 120km/h\cite{China}). $a_{max,accel}$ of $c_{1}$ and $c_{2}$ is $8.382m/s^2$ which is the maximum acceleration of Tesla\cite{Tesla}. $a_{min,brake,correct}$ and $a_{min,brake}$ is $6.2m/s^2$, which is the maximum reasonably skilled driver's braking acceleration\cite{braking}. Based on the above parameters, we can derive $\rho$, the safety time of forward cameras, so as to obtain the maximum response time allowed of each forward camera.

%%%%%%%%%%%%%%%%%%%%%%%%%%%%%%%%%%%%%%%%%%%%%%%%%%%%%%%%%%%%%%%%%%%%%%%%%%%%%

%Based on the Responsibility-Sensitive Safety (RSS) safety model\cite{Save}, the safety time of each camera can be got. When the two vehicles are driving at opposite directions, RSS proposes the formal mathematical Equation Lemma3 that identify the relationship between the minimal safe distances and processing time. Through setting the minimal safe distances to the max distance of each camera, and using the parameters (1) the maximum velocity allowed in different areas; (2) the maximum acceleration of each vehicle, for instance $8.382m/s^2$ of Tesla\cite{Tesla}; (3) the maximum reasonably skilled driver's braking - $6.2m/s^2$\cite{braking}, we can get the safety time of forward cameras based on Lemma3, so as to obtain the maximum response time allowed in automated vehicles. 

For the rear and side cameras, their safety time can be computed through Equation(\ref{safedis}) like forward cameras. To be noted that: (1) reversing will not be considered on the highway; (2) the maximum velocity of turning is set to 50km/h\cite{turn}. In summary, different cameras have different safety time, thus the maximum response time allowed of different cameras are different. In this paper, we define the matching score (MS) to indicate the relationship between the response time and the safety time (maximum response time) of each camera. %The maximum velocity allowed in UA, UHA and HA is 60km/h, 80km/h and 120km/h\cite{China} respectively.

%%%%%%%%%%%%%%%%%%%%%%%%important-Do not delete%%%%%%%%%%%%%%%%%%%%%%%%%%%%%%
%For the rear cameras, if the vehicle is reversing, and at the same time the obstacle is moving forward, thus the detection time of the image got by the rear camera should be the shortest. This shortest time is the safety time of this rear camera, and we can get it through Equation Lemma3 (we will not consider reversing on the highway).

%For the side cameras, if the vehicle and an obstacle will apply lateral acceleration toward each other, the detection time of the image got by the related side camera should be the shortest. This shortest time is the safety time of this side camera, and we can also get it through Equation Lemma3. Moreover, except for the maximum velocity always set to 50km/h\cite{turn}, all other parameters in the Equation Lemma3 are the same as the forward cameras. 
%%%%%%%%%%%%%%%%%%%%%%%%%%%%%%%%%%%%%%%%%%%%%%%%%%%%%%%%%%%%%%%%%%%%%%%%%%

\begin{figure}  
\centering
\includegraphics[width=3.3in,height=1.2in]{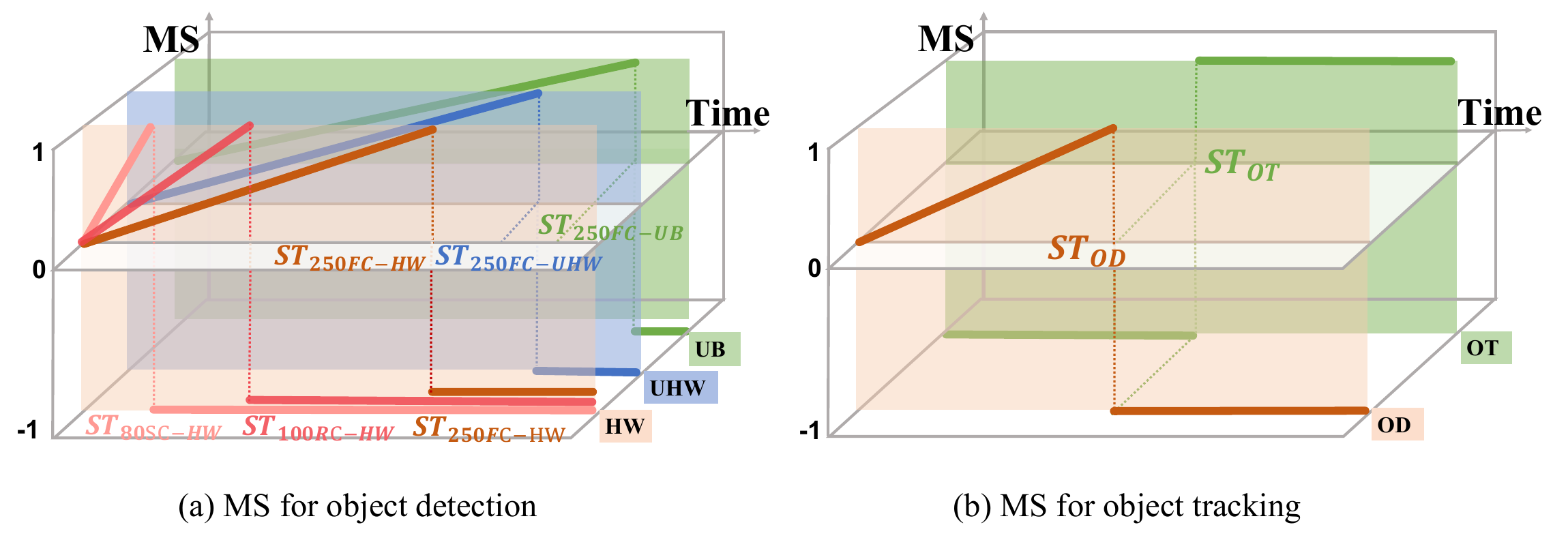}
\vspace{-2em} 
\captionsetup{font={small}} 
\caption{\textbf{MS in automated vehicles.} \label{TMS}}
\end{figure}

In Figure~\ref{TMS}(a), the horizontal axis represents the object detection tasks' response time for each camera, and the vertical axis represents MS. First, we analyze the MS of the same camera in different driving scenarios. $ST_{250FC-HW}$, $ST_{250FC-UHW}$ and $ST_{250FC-UB}$ represent the safety time of a forward camera with maximum distance 250 meters in HW, UHW and UB. We define [$0-ST_{250FC-HW}$],[$0-ST_{250FC-UHW}$] and [$0-ST_{250FC-UB}$] as accepted time (ACTime) regions, while treating [$ST_{250FC-HW}-\infty$],[$ST_{250FC-UHW}-\infty$] and [$ST_{250FC-UB}-\infty$] as unaccepted time (UACTime) zones. If a response time for a task lies in the ACTime region, its MS grows linearly as the time increases. This is because the energy consumption of the hardware would reduce as the execution time increases while safety time is guaranteed in this region \cite{song}. In the UACTime zone, the MS plummets to -1 due to the unacceptability of the response time. Furthermore, because the maximum velocity limit of the UB, UHW and HW is gradually increased, $ST_{250FC-UB}$, $ST_{250FC-UHW}$, and $ST_{250FC-HW}$ are gradually reduced accordingly.

Next, we introduce the MS for different cameras in the same driving scenario. As shown in Figure~\ref{TMS}(a), $ST_{250FC-HW}$, $ST_{100RC-HW}$ and $ST_{80SC-HW}$ represent the safety time of forward camera, rear camera and side camera with a maximum distance of 250, 80 and 100 meters respectively in HW. [$0-ST_{250FC-HW}$],[$0-ST_{80SC-HW}$] and [$0-ST_{100RC-HW}$] are ACTime regions, while [$ST_{250FC-HW}-\infty$], [$ST_{80SC-HW}-\infty$] and [$ST_{100RC-HW}-\infty$] are UACTime zones. The trend of MS for these three cameras in ACTime and UACTime are the same as above. 
%\footnote{\noindent We found some cameras have a invalid safety time by used Equation(\ref{safedis}) when the vehicle is driving at high velocity. At this time, we replace the safety time of this camera with other camera in the same direction which has the longer max distance, such as using FCS-250 to replace FCS-150 in HA.}. 

\noindent\textbf{Matching Score - Object Tracking}  
This section will introduce cameras' MS when its task type is object tracking. In the Figure~\ref{TMS}(b), $ST_{OD}$ and $ST_{OT}$ is the safety time of the same camera when its task type is object detection (DET) and object tracking (TRA) respectively. In autonomous driving system, TRA follows DET to predict the trajectories of moving objects\cite{TAIAD}, which indicates that TRA is processed after DET for the same image.
%, and is less sensitive to the response time. 
Therefore, $ST_{OT}$ should not be less than $ST_{OD}$, and we set $ST_{OT}$ equals to $ST_{OD}$ here. In the Figure~\ref{TMS}(b), [$0-ST_{OT}$] is ACTime, and [$ST_{OT}-\infty$] is UACTime. When TRA's response time of the current camera is in ACTime, MS is always -1, otherwise MS is 1.

%In the Figure~\ref{TMS}(b), the OD indicates the MS of object detection for a camera, and OT means the MS of object tracking for the same camera in the same area. 

\subsection{Global State Value}
To evaluate the overall performance of HMAI, we consider energy consumption $E$, runtime $T$ and resource utilization balance rate $R\_Balance$. $R\_Balance$ means the balance of resource utilization in HMAI, thus the higher $R\_Balance$, the less idle accelerators in HMAI at every moment. Whenever HMAI completes processing a task, these three values change accordingly. As the energy consumption of HMAI is expected to be as small as possible, the shorter the running time and the better the resource utilization balance rate to be, we define the Global State Value as $Gvalue=(-E-T+R\_Balance)/3$ (after normalization). 
\vspace{-0.5em} 
\section{FlexAI-A Task Scheduling Engine}

In the autonomous driving system, the dynamic environment can generate a massive amount of tasks, while the hardware resources are limited. Thus based on the metrics in criteria, how to designate tasks to different accelerators in HMAI needs to be carefully designed.

We use a real case to show the necessity of scheduling. Consider that when 30 cameras in a vehicle work once, then 30 frames will be generated simultaneously, thus we assume there will be 30 SSD tasks to process. We can not just allocate the same task to its best-fit accelerator because this will hurt the resource utilization of HMAI and overwhelm the chosen accelerator. Therefore, future driving automation platform needs an efficient task scheduling mechanism to trade-off among execution time, energy consumption, resource utilization and matching score.

The scheduling problem faced by HMAI is NP-complete. Conventional algorithms used to solve it can be classified into two groups, the heuristic-based and guided random-search-based algorithms\cite{Haluk}.
As for heuristic-based algorithms, \cite{ATA} proposed the Adaptive Task-partitioning Algorithm (ATA) to find out the scheduling policy of a task to consume as little energy as possible while guaranteeing the latency. The Min-min algorithm is considered as optimal in \cite{11}.
However, this algorithm can only consider the best hardware for each task while neglecting the global performance of HMAI. 
%As for heuristic-based algorithms, w-rand\cite{StarPU} guides the assignment of tasks by minimizing the energy and latency of hardware. Eleven heuristic algorithms are listed in \cite{11}, which points out that Min-min algorithm is optimal. However, this algorithm can only consider the best hardware for each task while neglecting the global performance of HMAI. Therefore, metrics other than the time, energy consumption and MS of a single task are ignored using this algorithm. E Oh et al. proposed the Adaptive Task-partitioning Algorithm (ATA)\cite{ATA}, of which the objective is to find out the scheduling policy of a CNN task to consume as little energy as possible while guaranteeing the latency limit of the task. Two dimensions---energy consumption and user demand for task execution time---are considered, but the hardware system in it only consists of multiple sets of identical hardware, and each task's execution time on each hardware needs to be obtained before the task queue starts being scheduled. The scheduling strategy proposed in\cite{EDP} produce a scheduling strategy by minimizing the Energy-Delay Product (EDP). In this case, only one dimension of the problem, i.e. the execution time of the task, is considered.

Genetic algorithms (GAs)\cite{hou,shroff,correa} and simulated annealing (SA)\cite{SA,SA2} are the most popular and widely used techniques for task scheduling problems in guided random-search-based algorithms. However, a fitness equation in GA and a cost function in SA are needed to select the optimal strategy for current tasks, thus the global performance like resource utilization of HMAI can't be taken into account.

In this paper, we propose FlexAI, a learning-based task scheduler to resolve the scheduling issues in driving automation system. Table~\ref{metrics} compares our work with other algorithms with respect to the coverage of the metrics proposed in Section 6.
%gives a comparison of these metrics for FlexAI and other algorithms. 
Specifically, to perceive the global performance in HMAI, we will use deep reinforcement learning (RL)\cite{RLO} in FlexAI as the scheduling algorithm introduced in Section 7.1. In Section 7.2, we will introduce how to use scheduling metrics to get the reward in RL.

\begin{table}[t]
\renewcommand\arraystretch{0.8}
\scriptsize
\centering
\begin{tabular}{|p{0.7cm}<{\centering}|p{0.3cm}<{\centering}|p{1cm}<{\centering}|p{0.3cm}<{\centering}|p{0.9cm}<{\centering}|p{0.6cm}<{\centering}|p{0.3cm}<{\centering}|p{0.7cm}<{\centering}|}
\hline
\multirow{2}{*}{\textbf{Metrics}} & \multicolumn{4}{c|}{\textbf{Heuristic}} & \multicolumn{2}{c|}{\textbf{Random-search}} & \multirow{2}{*}{\textbf{FlexAI}} \\ \cline{2-7}

& \textbf{EDP}\cite{EDP}  & \textbf{Min-Min}\cite{11}  & \textbf{ATA}\cite{ATA} & \textbf{W-rand}\cite{StarPU}  &\textbf{ GA}\cite{Haluk}      & \textbf{SA}\cite{SA2} &  \\ \hline

\textbf{Time} & \checkmark     & \checkmark  & $\times$  & \checkmark &\checkmark & \checkmark                  & \checkmark \\ \hline

\textbf{Energy}  & \checkmark & \checkmark & \checkmark    & \checkmark        &\checkmark                   &\checkmark  & \checkmark                       \\ \hline

\textbf{Resrc}    & $\times$ & $\times$ &$\times$ & $\times$ &$\times$ &$\times$ & \checkmark                       \\ \hline

\textbf{MS}  & $\times$& \checkmark& \checkmark  & $\times$ &\checkmark & \checkmark& \checkmark    \\ \hline
\end{tabular}
  \vspace{-1em} 
\captionsetup{font={small}}
\caption{ \textbf{The metrics of some algorithms and FlexAI.}
\label{metrics}}
\end{table}

\subsection{How the RL Agent Works}
We propose a reinforcement learning (RL)-based algorithm for task scheduling on the HMAI. A RL agent can learn strategy by interacting with the environment without any supervision. In each episode of learning, the agent can provide decision-making policies according to the current environment (HMAI) and the long-term objective. This is done by receiving feedback in form of a reward from the environment.

Assuming there are $N$ CNN accelerators \{$H_{1}$, $H_{2}$ ... $H_{N}$\} in the HMAI, and there will be $M$ tasks \{$A_{1}$, $A_{2}$ ... $A_{M}$\} coming in sequence, of which each is a CNN-based task like object detection based on YOLO or SSD, object tracking based on GOTURN. Then, the proposed RL algorithm generates scheduling strategy P = \{$p_{1}$, $p_{2}$ ... $p_{M}$\}, each of which indicates the task $A_{i}$ will be scheduled to $H_{j}$ under the guidance of $p_{i}$. When $A_{i}$ is executed, the metrics of HMAI will be updated accordingly, and the difference between the updated value and the previous value is denoted by reward $r_{i}$, thus the corresponding reward set is \{$r_{1}$, $r_{2}$ ... $r_{M}$\}.

%Note that the reward is generated by our estimation model during training, while is directly collected from HMAI in the inference. If the task types(YOLO, SSD, GOTURN in this paper) and task frequency do not change, the trained-well RL can be used all the time in automated vehicles.

%In this section, we propose deep reinforcement learning (RL) algorithm for task scheduling in the HMAI. RL can be performed without any supervision. After each episode, the RL agent will produce some results of interest, and the reward used to evaluate and update the RL agent will be obtained from environment. In FlexAI, as shown in Figure~\ref{FlexAI}, RL agent is used to produce scheduling policy for tasks in driving automation system. During training, the reward (with regard to metrics) is generated by the estimation model, and at inference, it is directly provided by HMAI.

\begin{figure}  
\centering
\includegraphics[width=3.2in,height=1.2in]{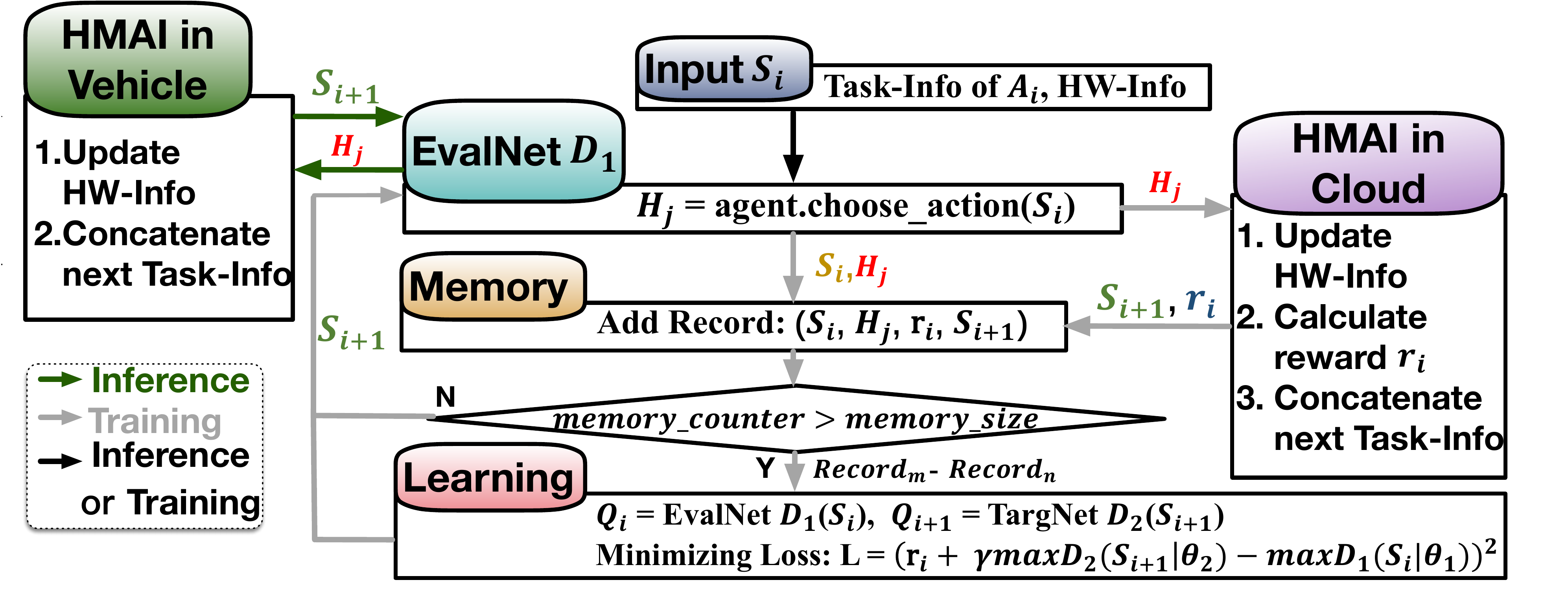}
 \vspace{-1em}
 \captionsetup{font={small}}
\caption{\textbf{The work flow of the RL scheduling agent. }\label{DQN}}
\end{figure}

The RL agent input contains three parts when training: (1) Task-Info including three parameters: \textbf{Amount}: the computation amount when task is processed; \textbf{LayerNum}: the CNN layer's number in the task; \textbf{safety time}: described in the Section 6.1. (2) HW-Info: the current information of all accelerators in the HMAI (the parameters of HW-Info will be described in Section 7.2). (3) Reward, in inference, there is no reward because the network doesn't need to update. 

%we will discuss the reason for choosing these parameters which constitutes LayerParam in Section 6.3

In this paper, we will use DQN to learn scheduling strategy from episodic job queues. \cite{ACR} divided the RL algorithm into three categories: critic-only (e.g. DQN~\cite{DQN}), actor-only (e.g. Policy gradient~\cite{PolicyG}) and actor-critic (e.g. DDPG~\cite{DDPG}). When the scheduling strategy of a single task is generated, the critic-only algorithm can be trained once directly. However, the actor-only algorithm can only be trained after the scheduling strategies of all tasks in a task queue are generated. Since the number of tasks in each task queue in autonomous driving system is extremely large (up to 30,000 introduced in Section 8.3), to reduce training time, we will not choose actor-only category. Furthermore, although actor-critic can be trained in the same way as critic-only, due to its high computation complexity, we will choose DQN that falls into the critic-only category.

In our method, the two networks are denoted by EvalNet $D_1$ with the parameter $\theta_1$ and TargNet $D_2$ with the parameter $\theta_2$. EvalNet is used to generate the scheduling policy of the current task, while TargNet is used to update the parameters of the EvalNet. These two networks are consisting of two fully connected layers, and their input $S_{i}$ is Task-Info and HW-Info of task $A_{i}$. The output of these two networks is a group of Q values $Q_{j}$. $Q_{j}$ is the cumulative value of the reward: $Q_{j}=\sum_{^{n=i}}^{M}r_{n}$, which is generated after the task $A_{i}$ was executed on the $H_{j}$. After obtaining $N$ Q values, EvalNet or TargNet will choose $H_j$ which attains the maximum Q value after $A_i$ is executed. The choice $H_j$ of EvalNet or TargNet is a scheduling strategy $p_{i}$ which guides to schedule task $A_{i}$.

Figure~\ref{DQN} shows the working process of our RL scheduling agent. First, EvalNet $D_1$ will generate scheduling strategy for the input $S_{i}$, and use it to allocate task $A_{i}$ to $H_{j}$. Then, in training, (1) HMAI in cloud uses $H_{j}$ to update HW-Info, and calculates reward $r_{i}$. Next, HW-Info will combine with Task-Info of next task $A_{i+1}$ to generate $S_{i+1}$; (2) the record ($S_{i}, H_{j}, r{i}, S_{i+1}$) is saved in memory, and if the total amount of records in the memory is greater than the memory size at this time, the RL agent will use $record_{m}$ - $record_{n}$ to start learning. (3) in learning, as for $record_{i}$, Eval\_net $D_1$ will use $S_{i}$ to generate $Q_{i}$, and TargNet $D_2$ will use $S_{i+1}$ to generate $Q_{i+1}$. Then $\theta_1$ is updated by minimizing the loss:
$L = {(y_{i}-maxD_1(s_i |\theta_1))}^2$, where $y_{i} = r_{i} + \gamma maxD_2(s_{i+1} |\theta_2)$. The parameter $\theta_2$ in $D_2$ will be copied directly from $D_1$ every fixed time. In inference, HMAI in vehicle uses $H_{j}$ to update HW-Info, and then $S_{i+1}$ is sent to the EvalNet $D_1$ directly.

\subsection{The Way to Get Reward}

In this section, we will introduce how to use scheduling metrics to get the reward in RL. Suppose that the information (Info) of $H_i$ is ($E_i, T_i, R\_Balance_i, MS_i$), and we use Info for each accelerator in HMAI to constitute the \textbf{HW-Info}. After $A_j$ is scheduled to $H_i$, the energy consumption, time, MS and resources utilization balance rate of processing $A_j$ are denoted by $e_j$, $t_j$, $ms_j$ and $r_j$. Thus, for $H_i$, (1)$E_i+=e_j$; (2)$T_i+=t_j$; (3)$MS_i+=ms_j$; (4)$R\_Balance_i=\frac{r_j+R\_Balance_i}{num}$ (num is the number of tasks has been executed in $H_i$). Until now, the energy consumed in each accelerator is $\{E_{1},E_{2}...E_{N}\}$ respectively, the total time is $\{T_{1},T_{2}...T_{N}\}$, the resource utilization is $\{R\_Balance_{1},R\_Balance_{2}...R\_Balance_{N}\}$, and the sum of the $MS$ in each accelerator is $\{MS_{1},MS_{2}...MS_{N}\}$. Then for HMAI, (1) $E=\sum_{i=1}^{N}E_i$; (2) $R\_Balance=\frac{1}{N}\sum_{i=1}^{N}{R\_Balance_i}$; (3) $MS=\sum_{i=1}^{N}MS_i$; (4) $T=\max\{T_1,T_2...T_N$\}. 

Now if there are currently $M-1$ tasks scheduled to HMAI, at this moment, the HMAI has $E$, $T$, $R\_Balance$, and $MS$. When the $Mth$ task is executed, the four values will be updated to $E_{new}$, $T_{new}$, $R\_Balance_{new}$, and $MS_{new}$. Then, after processing $Mth$ tasks \textbf{reward} is given by $(-E_{new} - T_{new} + R\_Balance_{new})/3-(-E - T + R\_Balance) /3+ MS_{new} - MS = Gvalue_{new} - Gvalue + MS_{new} - MS$.

%All these quantities are used to indicate the status of each core in the HMAI after a task has been executed.

\section{Evaluation}
%%%%%%%%%%%%%1%%%%%%%%%%%%%%%
\begin{table}[t]
\renewcommand\arraystretch{1}
  \centering
  \scriptsize
  \begin{tabular}{ p{2.3cm}<{\centering}p{5.3cm}<{\centering} }
    \hline 
    \textbf{Parameter} & \textbf{Description} \\
    \hline
    \multicolumn{1}{r}{\multirow{3}{*}{Camera\_HZ(A, S, C)}} & A includes UB, UHW, HW\\
    \cline{2-2} 
     \multicolumn{1}{r}{}   & S includes Go straight, Turn and Reverse\\
    \cline{2-2} 
    \multicolumn{1}{r}{}   & C includes  FC, FLSC,RLSC, FRSC, RRSC and RC\\ 
    \hline
    MaxTimes\_Turn(A)& A includes UB, UHW, HW\\ 
    \hline
    MaxTimes\_Reverse(A)& A includes UB, UHW, HW\\
     \hline
    MaxDuration\_Turn(A)& A includes UB, UHW, HW\\
     \hline
    MaxDuration\_Reverse(A)& A includes UB, UHW, HW\\
     \hline
    Velocity(A)& A includes UB, UHW, HW\\
    \hline
    \multirow{2}{*}{Safety\_Time(A, C)}  & A includes UB, UHW, HW  \\
    \cline{2-2} 
    & C includes  FC, FLSC,RLSC, FRSC, RRSC and RC\\ 
    \hline
  \end{tabular}
    \vspace{-1em} 
    \captionsetup{font={small}}
    \caption{\textbf{The parameters in Autonomous Driving Environment.} \label{pEn}}
\vspace{-2em}
\end{table}
\subsection{The Dynamic Driving Environments}
To simulate a variety of vehicle driving areas and scenarios, we define several parameters (listed in Table~\ref{pEn}), which characterize dynamic driving environments. Note that when the parameter A (urban areas (UB), undivided-highways (UHW), highways (HW)) changes, the frequency of cameras (Camera\_HZ), the maximum number of turning ((MaxTimes\_Turn) and reversing (MaxTimes\_Reverse), the longest duration of turning (MaxDuration\_Turn) and reversing (MaxDuration\_Reverse), the speed of the vehicle (Velocity), and the safety time of cameras (Safety\_Time) will all change. Moreover, when the parameter C (FC, FLSC, RLSC, FRSC, RRSC and RC) changes, Camera\_HZ and Safety\_Time will vary. Similarly, when the parameter S (Go straight, Turn and Reverse) changes, Camera\_HZ will change also. Table~\ref{ccids} lists the 
number of cameras in each type.

\makeatletter\def\@captype{table}\makeatother
\begin{minipage}{.22\textwidth}
\vspace{-1em} 
\centering
\vspace{+2.5em} 
\scriptsize
\begin{tabular}{p{1.8cm}<{\centering}p{0.5cm}<{\centering}}
\hline
\multicolumn{2}{c}{\textbf{Parameter Setting}} \\ \hline
MaxTime\_Turn       & 10                \\
MaxTimes\_Reverse   & 10                \\
MaxDuration\_Turn   & 10                \\
MaxDuration\_Reverse& 20                \\ \hline
\multicolumn{2}{c}{\textbf{Current Setting}}   \\ \hline
Turn Times          & 2                 \\
Reverse Times       & 1                 \\
Turn Duration       & 3s,4s             \\
Reverse Duration    & 2s                 \\ \hline  
  \end{tabular}
%\vspace{+1em} 
 \captionsetup{font={small}} 
\vspace{+0.5em}  
\caption{\textbf{The parameters.} \label{insPA}}
\vspace{+2em}  
\end{minipage}
\makeatletter\def\@captype{figure}\makeatother
\begin{minipage}{.20\textwidth}
\centering
\vspace{-1em} 
\includegraphics[width=1.2in,height=1.4in]{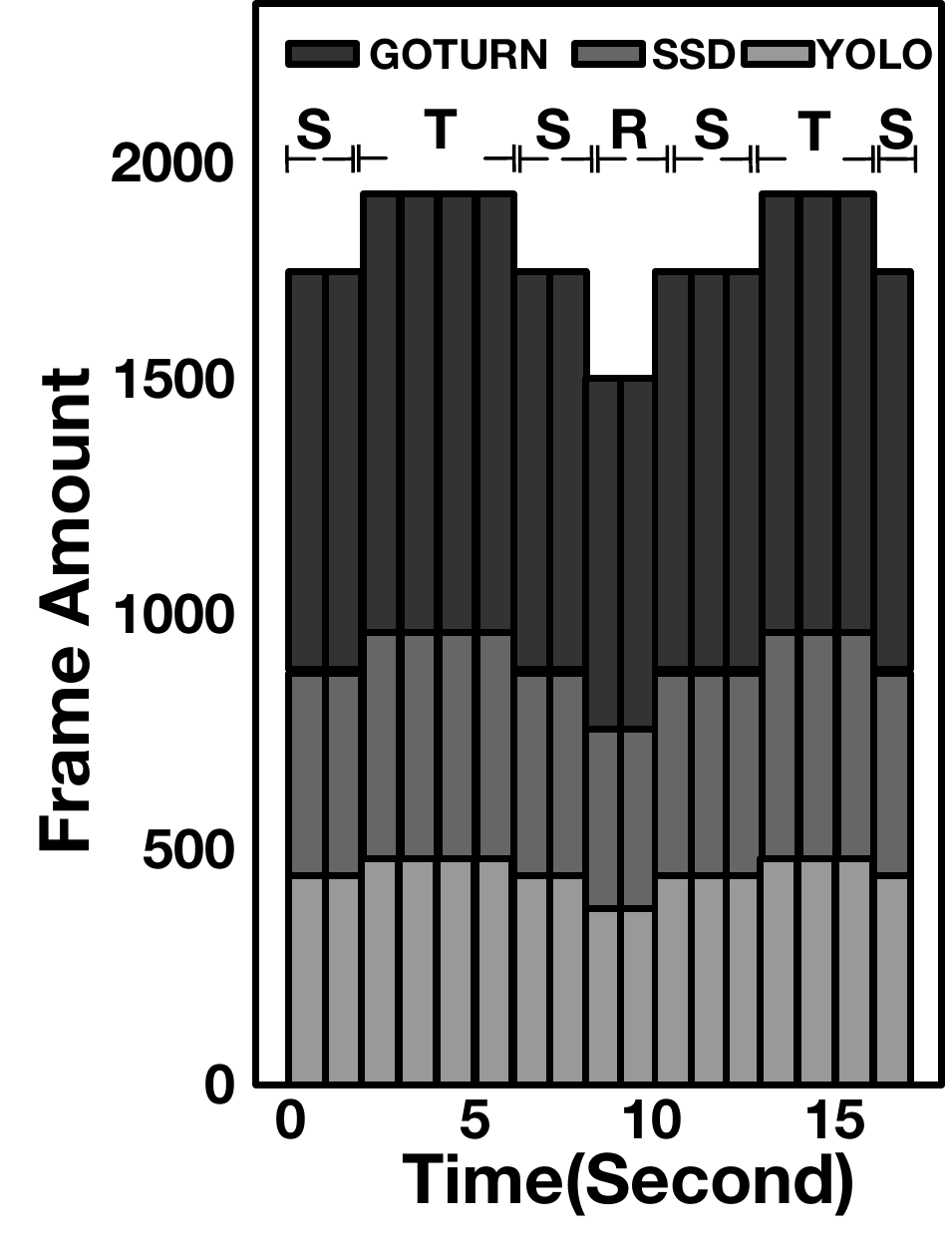}
\vspace{-0.5em} 
\captionsetup{font={small}}
\caption{\textbf{The task queue. }\label{120m}}
\end{minipage}
 \vspace{-1em}

\noindent\textbf{Task Queue.}
In this experiment, we use images in the KITTI object tracking 2D dataset~\cite{Geiger2012CVPR} as tasks in our task queue. This object tracking dataset consists of multiple sequences, and the images in each sequence are the continuous outputs for one camera in our vehicle. Based on the parameters in Table~\ref{pEn}, we create different driving routes with various driving distances, and all tasks generated by the vehicle during the route form one task queue. In addition to the dataset that constitutes the task queue, we also specify the task amount at different time in this task queue. Figure~\ref{120m} illustrate an example of a task queue when a vehicle has a 160m route in UB and its velocity is 20m/s. Parameters such as Camera\_HZ, Safety\_Time are derived from Figure~\ref{cfa}, Section 6.1 and Table~\ref{insPA}. As mentioned in Section 2, we alternately use YOLO and SSD to process the DET tasks for each camera, and use GOTURN to process TRA tasks, thus the task types in Figure~\ref{120m} are YOLO, SSD and GOTURN. In Figure~\ref{120m}, S, T and R indicate three scenarios: going straight, turning and reversing, and the start time and lasting time of each scenario is randomly determined.

%%%%%%%%%%%%%2%%%%%%%%%%%%%%%
\vspace{-0.5em}
\subsection{Performance Analysis for HMAI}
\noindent\textbf{The construction of HMAI.} In Figure~\ref{newhohe2} (b), we give the resource utilization rate of each platform in urban areas. Comparing the geometric mean of resource utilization rates in three scenarios of each platform, we find (4 SconvOD, 4 SconvIC, 3 MconvMC) is the best configuration. According to the same method, in the other two areas, above heterogeneous platform configuration can also achieve optimal resource utilization. Therefore, we choose to use (4 SconvOD, 4 SconvIC, 3 MconvMC) to construct our heterogeneous platform HMAI.

\noindent\textbf{Experimental Methodology.} The performance and energy of HMAI is measured by the following tools. For the performance evaluation, a customized cycle-accurate simulator was designed and implemented to measure execution time in number of cycles. This simulator models the microarchitectural behavior of each hardware module of our design. In addition, we use ARM1176 as the main control processor in the HMAI to do task scheduling.

For measuring area and power, we implemented a Verilog version of each hardware module, then synthesized it. We used the Synopsys Design Compiler with the TSMC 12nm standard VT library for the synthesis, and estimated the power consumption using Synopsys PrimeTime PX. In addition, the design of SconvOD, SconvIC and MconvMC is based on ~\cite{NeuFlow, ShiDianNao,Origami}. And the SRAM is generated by Synopsys Memory Compiler and the interconnect bus is generated by Synopsys DesignWare AMBA IP.

%%%%%%%%%%%%%%%
\iffalse

\begin{figure}[t]
\centering
\includegraphics[width=2.4in,height=1in]{ISCA2019-latex-template/figure/WechatIMG1047.jpeg}
\vspace{-1em} 
\captionsetup{font={small}} 
\caption{\textbf{Latency and energy breakdown.}}\label{bdTE}
\vspace{-1em} 
\end{figure}

\noindent\textbf{Power/Area} Figure~\ref{bdTE} shows the power and area breakdown of HMAI. The total power and area of HMAI are 158 W and 132 mm2 respectively. sensor controller only cost X\% power as well as X\% area. SoC interconnect costs X\% power as well as X\% area. Meanwhile CPU costs X\% power as well as X\% area since it is only used to do task scheduling. Data SRAM costs X\% power as well as X\% area due to it size is 60MB. 
\fi

\noindent\textbf{Baseline.} To compare HMAI with state-of-the-art work, we evaluate HMAI with NVIDIA Tesla T4 GPU, which is mainly designed for AI inference workloads. In addition, in order to prove that heterogeneous platform is better than homogeneous platform, we compare the performance of HMAI with multiple homogeneous platforms. The homogeneous platforms compared here have been mentioned in Section 3.1, which are 13 SconvOD, 13 SconvIC and 12 MconvMC.

\noindent\textbf{The Parameters for Constructing Task Queue and Experimental  Methodology.} Here, we create different driving routes for urban area (UB) with various distances from 1km to 2km, and vehicle's velocity is set to 60km/h. In Experimental, first, 5 task queues are constructed, and then we use HMAI including FlexAI, NVIDIA Tesla T4 and three homogeneous platforms to process each task queue. 

\begin{figure}[t]  
\centering
\includegraphics[width=3.4in,height=1.3in]{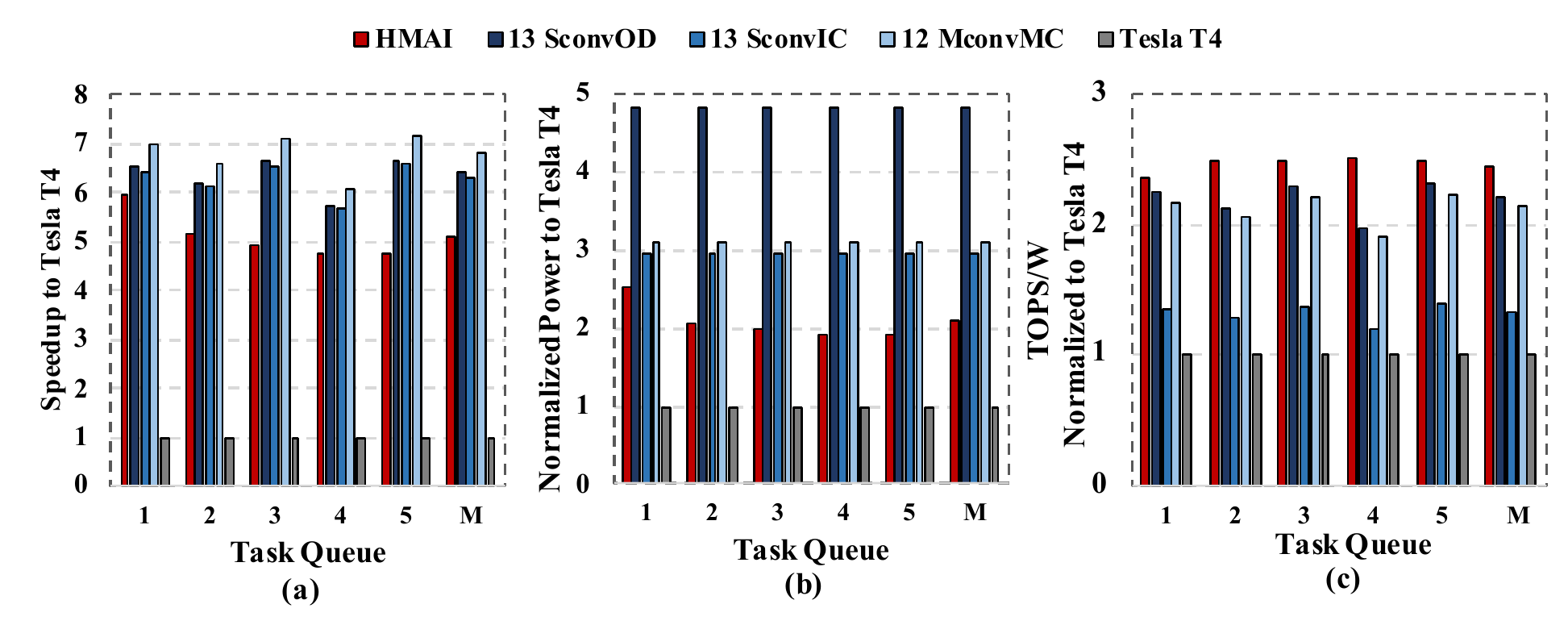}
\vspace{-2.5em}
\captionsetup{font={small}} 
\caption{\textbf{Comparison between HMAI and baselines. The last set of bars, labeled as M, indicate the geometric mean across all task queues.}}\label{HMAIho}
\vspace{-1em} 
\end{figure}

\noindent\textbf{Experimental Results.}
We first compare the performance speedup normalized to Tesla T4, as shown in Figure~\ref{HMAIho} (a). Overall, HMAI achieves 5× speedup over Tesla T4, since HMAI provides more sufficient computing units for autonomous driving performance requirements. Note that, different autonomous driving scenarios have different performance requirements (detail in Section 3.1). In order to satisfy all requirements, the computing resources in the homogeneous platforms will be redundancy in some scenarios, thus there will be a task which response time is much shorter than its safety time. Therefore in Figure~\ref{HMAIho} (a), all homogeneous platforms achieve higher speedup compared to HMAI. However, in various driving scenarios, by reasonably using the computing resources in heterogeneous HMAI, HMAI will have high resource utilization rate as shown in Figure~\ref{newhohe2} (b) and safe response time for each task.

Figure~\ref{HMAIho} (b) shows the normalized power of HMAI and the the homogeneous platforms, normalized to Tesla T4. HMAI reduces power by 57\%, 30\% and 33\% compared to three homogeneous platforms on average. The main reason is the reduction of a large number of redundancy computing resources. Since HMAI has much more computing resources than Tesla T4, it has about 2× power over Tesla T4. Theoretically, 5 Tesla T4 can have sufficient performance for autonomous driving, while the corresponding power will also be increased 5× than Tesla T4 itself. Here HMAI has only 2× power over Tesla T4, thus in Figure~\ref{HMAIho} (c), HMAI has higher TOPS/W than Tesla T4. In Figure~\ref{HMAIho} (c), HMAI also has higher TOPS/W than three homogeneous platforms due to its high resource utilization rate.

%As we proved in Section 4.1, resource utilization rate $R$ of HMAI always is the best compared with all the homogeneous platforms. As shown in Figure~\ref{HMAIho} (b), HMAI can maximally improve the resource utilization rate by 20\%, 15\% and 19\% compared to 13 SconvOD, 13 SconvIC, and 12 MconvMC. For providing sufficient hardware resources for all scenarios, in some cases, the resources are redundant in the homogeneous platforms. However, with the help of FlexAI, HMAI can meet the performance requirements of all scenarios with minimal resources. By using the FlexAI as well, in Figure~\ref{HMAIho}, MS and $E$ in HMAI are always best compared with all the homogeneous platforms. In Figure~\ref{HMAIho} (c), HMAI can maximally improve the MS by 63\%, 38\% and 105\% compared to 13 SconvOD, 13 SconvIC, and 12 MconvMC.  The range of MS is from -1 to 1, and we hope that the MS is as large as possible. In Figure~\ref{HMAIho} (d), HMAI can maximally reduce the power consumption by 4\%, 10\% and 7\% compared to 13 SconvOD, 13 SconvIC, and 12 MconvMC.%MS defined in Section 5.1 indicates the relationship between the response time of HMAI and the required time of different cameras in vehicles. T

%For the execution time, although in HMAI it is not the smallest, the MS of HMAI always is the best. As for autonomous driving, it is more important to make the task execution time meet the requirements than to make the task be executed in the shortest time, so although $T$ in HMAI is not the smallest, it has less impact in autonomous driving.

%%%%%%%%%%%%%%%%%%%%%%%%%%%%%%%%import%%%%%%%%%%%%%%%%%%%%%

%%%%%%%%%%%%%3%%%%%%%%%%%%%%%
\vspace{-0.5em}
\subsection{Performance Analysis for FlexAI}
In this section, we set up experiments to compare our RL-based FlexAI with other state-of-the-art schedulers.

\noindent\textbf{Baseline.} We use Min-Min, ATA in heuristics, GA, SA in guided random search techniques, as well as the unscheduled worse case as our baselines.

\noindent\textbf{The Parameters for Constructing Task Queue.} We create different driving routes for urban area (UB), undivided-highways (UHW), and highways (HW) with various distances from 1km to 2km, and velocity is set to 60km/h, 80km/h and 120km/h respectively\cite{China}.

\noindent\textbf{Training.}
The DQN used in FlexAI agent includes two networks with exactly the same structure but different updating paces. Each network is comprised of two fully connected layers, and a softmax layer. The number of neurons of the fully connected layers are 256 and 64 with ReLU non-linearity. We train three RL agents for UB, UHW, and HW respectively. Each agent is trained on the NVIDIA TITAN-XP with 1000 episodes, and each episode includes one task queue. The learning rate for training the EvalNet is 0.01.

\begin{figure}[t]
\centering
\includegraphics[width=3in,height=1.3in]{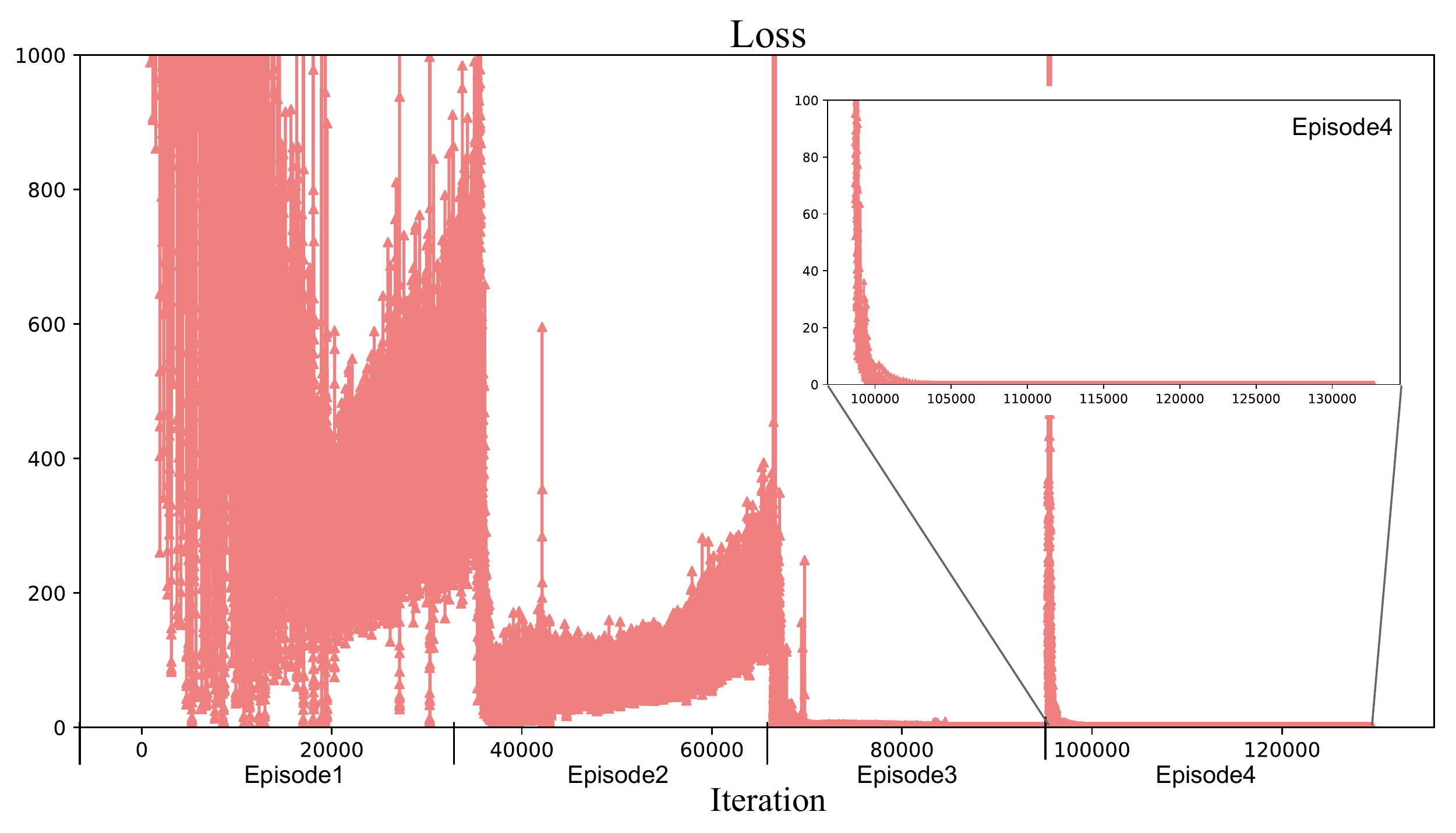}
\vspace{-1em} 
\captionsetup{font={small}} 
\caption{\textbf{Loss curve of RL agent. The threshold of the y-axis is 1000.}\label{loss}}
\vspace{-2em} 
\end{figure}

\noindent\textbf{Training Loss Curve.}
Figure~\ref{loss} shows the training loss curve of FlexAI RL agent in urban area. Each iteration represents one task, and each episode contains one task queue. As all object detection and object tracking tasks generated by a vehicle in a 1km - 2km route will form one task queue, each episode will contain up to 30,000 tasks. In Figure~\ref{loss}, the loss of the second episode gradually stabilizes after 10,000 iterations, while in the third and fourth episodes, except for the loss of the initial 2,000 iterations, the subsequent loss gradually tends to 0. The reason for that is the composition of tasks in each episode are very similar, thus the network trained in prior episodes will be applicable to subsequent episodes. This also further illustrates that if the task types do not change, the well-trained RL agent can be used all the time in automated vehicles.

\noindent\textbf{Experimental Methodology.}
For each area, first we use well-trained agent of FlexAI and each baseline to generate the scheduling strategy for 5 task queues. Next we compare the metrics between FlexAI and baselines for each task queue.

%\iffalse
%\fi

\begin{figure*}  
\centering
\vspace{-1.5em}
\includegraphics[width=6.8in,height=2.2in]{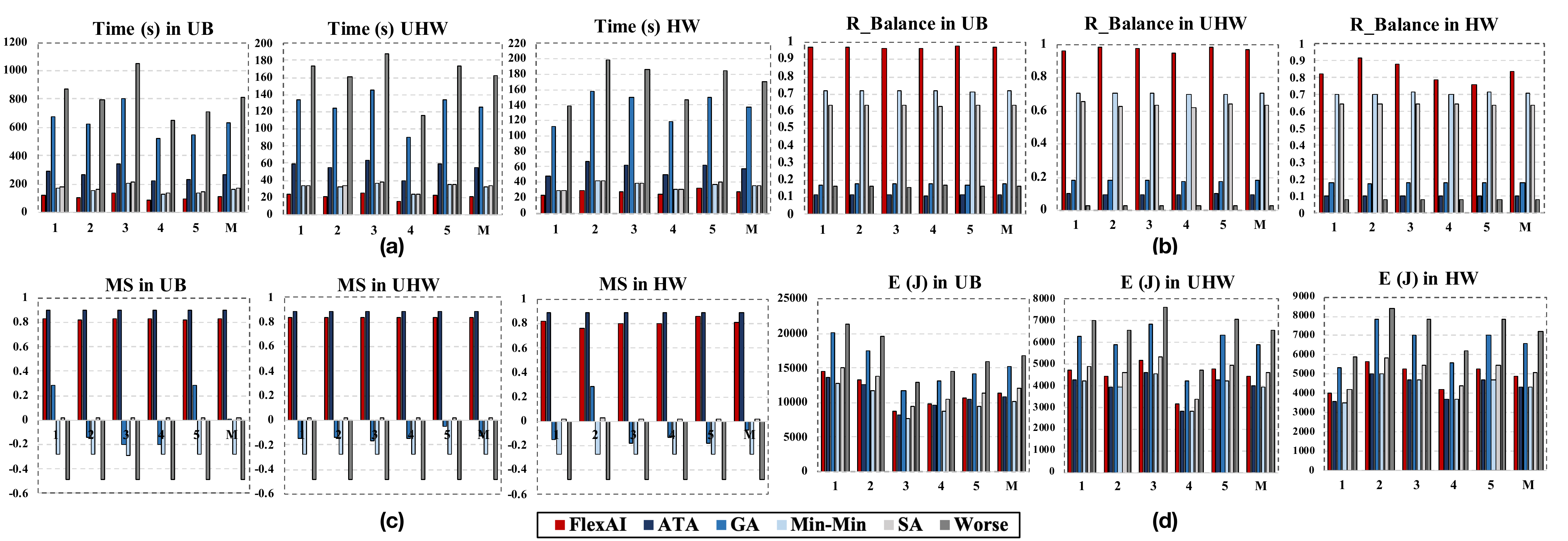}
\vspace{-1em}
\captionsetup{font={small}} 
\caption{\textbf{Comparison between FlexAI and baselines. UB, UHW and HW means urban areas, undivided-highways and highways. M is the geometric mean of 5 task queues.}}\label{SEC}
\vspace{-1.5em} 
\end{figure*}

\noindent\textbf{Experimental Results.}The time in Figure~\ref{SEC} includes three parts: (1) scheduling strategy runtime in CPU, (2) task waiting time and (3) task execution time. In Figure~\ref{SEC}(a), FlexAI can maximally reduce the time by 60\%, 88\%, 33\%, 36\% and 87\% compared to ATA, GA, Min-Min, SA and worse case in urban area. And for the geometric mean, FlexAI decreases at most 87\%. The reason for this is FlexAI can effectively reduce the task waiting time, and more details can be found in Section 8.4. For Min-Min, SA and ATA, they perform close to the FlexAI does since they consider execution time when scheduling. However, due to the fact that GA's performance is affected by the selection of the initial population, its time is much large than FlexAI. To summary, FlexAI can always achieve the minimum time in three areas, and this will ensure the safety of autonomous driving.
%although the time has been considered as well. 

As shown in Figure~\ref{SEC}(b), the $R\_Balance$ of FlexAI has been maximally improved by 837\%, 957\%, 62\%, 55\% and 960\% compared to ATA, GA, Min-Min, SA and worse case in urban area. As for the geometric mean in all areas, $R\_Balance$ in FlexAI is always the best. This is because among all scheduling strategies, only FlexAI considers the balance of resource utilization. The situation of $R\_Balance$ in the other two areas is the same as the urban area. In HMAI, by increasing $R\_Balance$, the task waiting time can be reduced, and at the same time it can decrease the waste of the hardware resources and improve the vehicle’s endurance.

In Figure~\ref{SEC}(c), MS in FlexAI is larger than that in GA, Min-min, SA and worse case (up to 1.02, 1.12, 0.83, 1.32), however smaller than that in ATA in the urban area. The reason is ATA is optimized towards MS, but FlexAI needs to tradeoff among four metrics. In addition, except for ATA and FlexAI, the other baselines' MS are always less than 0, which means there are many tasks in each task queue whose processing time are greater than the safety time. The situation of MS in the other two areas is the same as the urban area. In autonomous driving, higher MS represents better safety, and more discussion can be found in Section 8.4.

Figure~\ref{SEC}(d) shows the comparison of energy. Although FlexAI can achieve lower energy than GA, SA and worst case in all areas, it is slightly higher than the others. Some reasons are as follow. First, energy-performance tradeoff is common in accelerators. Moreover, FlexAI needs to consider $T$, $E$, $R\_Balance$ and MS at the same time, which makes tradeoff more difficult. 

%When an accelerator executes a task, it may take relatively least time, but the energy can not always be the smallest in the mean time. 

%%%%%%%%%%%%%4%%%%%%%%%%%%%%%
\vspace{-0.5em}
\begin{figure}  
\centering
\includegraphics[width=3.2in,height=0.8in]{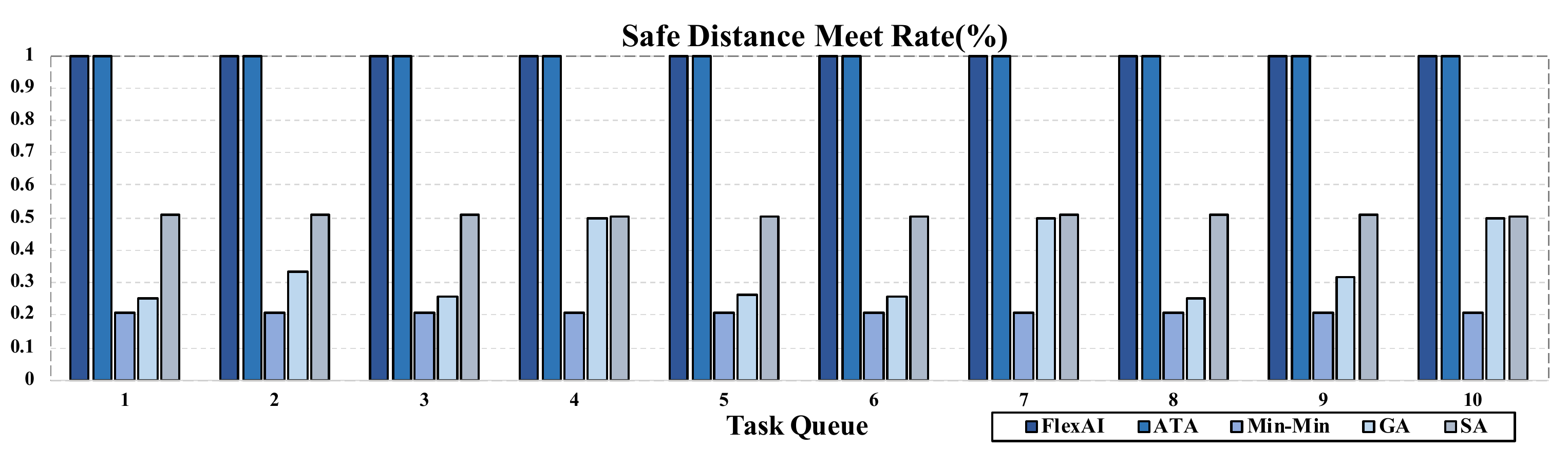}
\vspace{-1em}
\captionsetup{font={small}} 
\caption{\textbf{Safe distance meet rate (STMRate).}}\label{vMS}
\vspace{-1em} 
\end{figure}

\subsection{Autonomous Driving Metrics}
\noindent\textbf{Safety Time Meet Rate}
As we mentioned in Section 6.1, since each camera in a vehicle has a corresponding safety time, each task will have its safety time according to the camera that generated it. In this section, we define safety time meet rate (STMRate) to describe the proportion of tasks in a task queue whose processing time is less than its safety time. In Figure~\ref{vMS}, for each task queue, the STMRate of FlexAI is basically close to 100\%, which means the processing time of almost all tasks can ensure the driving safety. The reason for that is FlexAI considers MS when generating scheduling strategies, and MS indicates the relationship between the task processing time and task safety time. Here, since ATA is optimized towards MS, the STMRate of each task queue is also very high under ATA.

\begin{figure}[t]
 \centering
%\vspace{-0.5em}
\includegraphics[width=3.2in,height=1.7in]{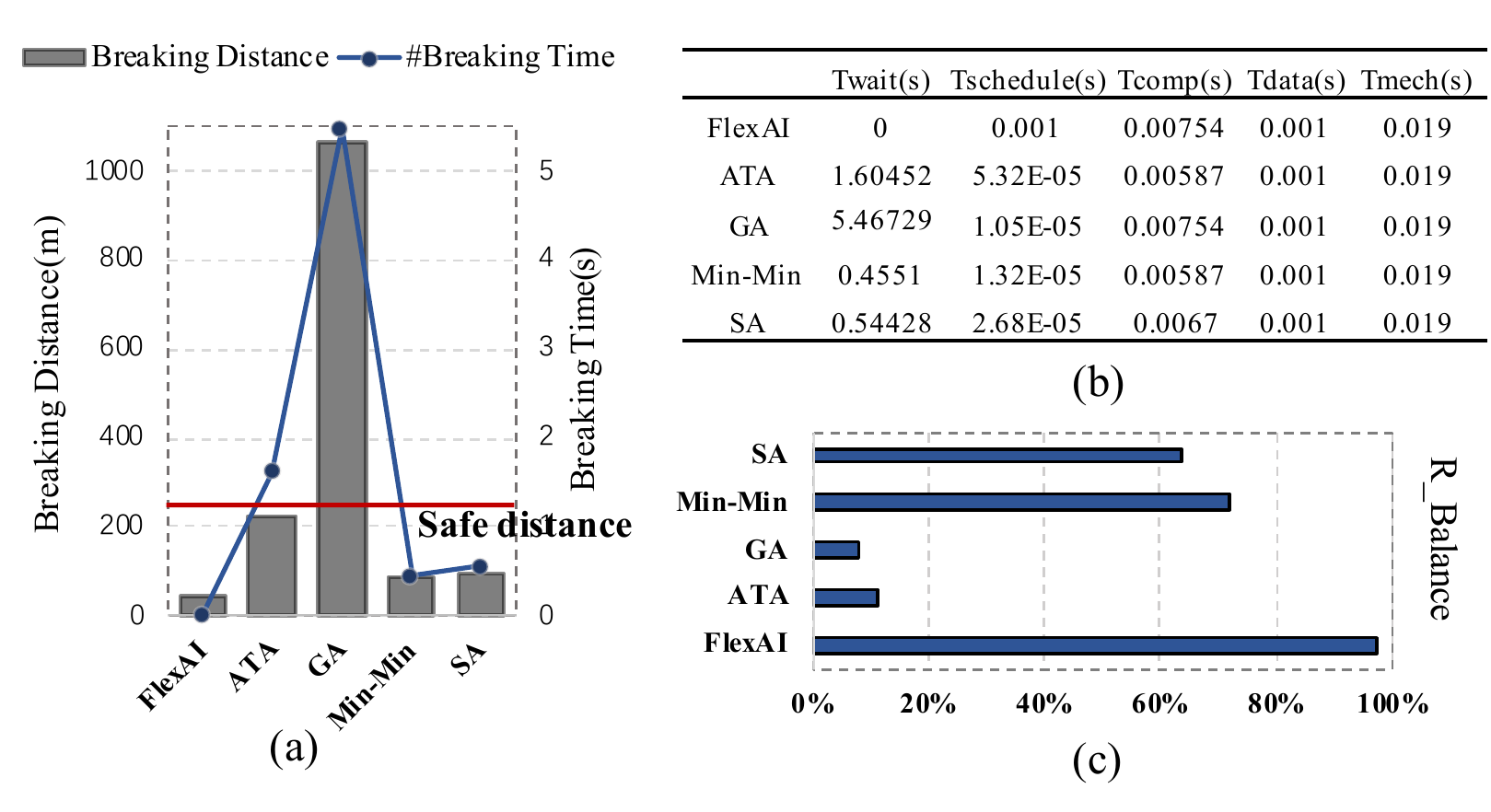}
\vspace{-1.5em}
\captionsetup{font={small}}
\caption{\textbf{(a)Breaking distance and total breaking time; (b)Total breaking time breakdown; (c)R\_Balance.}\label{bande}}
\vspace{-2em} 
\end{figure}

\noindent\textbf{Braking Distance.}
Here, we assume that after a vehicle moves 1km, its forward camera finds there is an object 250 meters away, so it needs to take braking immediately. The current velocity of the vehicle is 60 km/h, and the braking deceleration is 6.2 $m/s^2$. In Figure~\ref{bande} (a)(bar), the breaking distances under different schedulers are presented.

As shown in Figure~\ref{bande} (a)(bar), except for GA, the breaking distances of other schedulers are all less than the safe distance 250 m, and FlexAI has the smallest breaking distance 47.08 m. Since the breaking distance is strongly related to the braking time (breaking distance is calculated based on the Equation(\ref{safedis}) in Section 6), FlexAI has the smallest breaking time, as shown in Figure~\ref{bande} (a)(blue line).

The total breaking time breakdown of each scheduler is shown in Figure~\ref{bande} (b). $T_{wait}$ is the waiting time of the current task (the current task is used to detect the object that caused the braking); $T_{schedule}$ represents the runtime of each scheduler; $T_{compute}$ is the processing time of the current task in HMAI; $T_{data}$ is the time which is used to transmit the control commands to the vehicle’s actuator through the Controller Area Network (CAN) bus, which is 1 ms in this vehicle\cite{yu2020building}; $T_{mech}$ is the time that mechanical components of the vehicle takes to start reacting, which is 19 ms. In Figure~\ref{bande} (b), $T_{wait}$ in FlexAI is 0, while $T_{wait}$ is much larger than the time of other parts in other schedulers. Therefore, although $T_{schedule}$ and $T_{compute}$ of FlexAI are not the best, the total breaking time of FlexAI is the smallest.

Note that in our experiment, the vehicle generates a task queue from starting to braking. As shown in Figure~\ref{bande} (c), $R_{Balance}$ in FlexAI is the largest. It means under FlexAI, the resource utilization in HMAI is the most balanced, thus the number of idle accelerators in HMAI at every moment is the least as well. For instance, if task A has the fastest execution time in the accelerator A and accelerator A is currently busying, FlexAI will schedule task A to other idle accelerators to reduce its waiting time, while for Min-Min, task A will be waiting until accelerator A changes to idle. Therefore, for FlexAI, since it has the highest $R\_Balance$, its $T_{wait}$ in Figure~\ref{bande} (b) is the smallest. And under the smallest $T_{wait}$, FlexAI has the smallest braking distance in Figure~\ref{bande} (a)(bar).

\vspace{-0.5em}
\section{Related Work}
%For the related work of estimating mechanism, \cite{Eyeriss1} gives the method to estimate the transmission power consumption for different dataflows in accelerators, and the transmission paths of data are assumed and fixed in it. Besides, in \cite{Eyeriss1}, the design of various dataflows in the deep learning accelerator is summarized by the classification standard --- the way of data reuse in PE array. Based on the optimal storage structure and the size in the accelerator, the estimating model in \cite{MAESTRO} outputs the estimates of the accelerator's maximum performance. In contrast, the estimating model of accelerator proposed in this paper is mainly based on the propagation path of each data in real hardware. 
To the best of our knowledge, an emerging line of research has found that reinforcement learning proves to be effective in solving scheduling problems in various domains. \cite{UAV} uses RL to make real-time decision for the scheduling problem in flying mobile edge computing platform, and the reward of this RL agent includes the energy consumption of all user equipment. \cite{grid} proposes a multi-agent reinforcement learning approach for job scheduling in grid computing, of which the reward consists of the total execution time of all jobs. \cite{Cognitive} uses RL to deal with the radar resource management problem when the radar assigns limited time resource to a set of tasks, and the reward is comprised of the number of tasks delayed or dropped. For mobile-edge computing system, \cite{MEC5G} proposes a RL based task scheduling algorithm, with its reward involving the slowdown and average timeout period of tasks.

As the rewards of the algorithms aforementioned are designed for their specific scenarios, these ad-hoc formulations cannot be used to solve the scheduling problem in autonomous driving. In autonomous driving systems, each CNN-based task should be handled separately, and the reward should take into account not only the dynamic changes of the environment (HMAI), such as current resource utilization, but also the total energy consumption and the longest execution time among all accelerators. Furthermore, whether the current strategy meets the real-time requirements of the cameras in autonomous vehicles also matters. Therefore, it is desirable to develop RL-based scheduling algorithms for autonomous driving.

%For the related work of using machine learning (ML) to solve problems, a ML model is used in \cite{Related1} to predict the execution of the task in the system, including client satisfaction degree of each task and system power consumption. This learning method is supervised, which requires a huge amount of annotation. Additionally, this work predicts various conditions of task execution rather than scheduling strategies. In \cite{P1}, a deep reinforcement learning (RL) is used to implement task scheduling on cluster. It assumes as apriori the hardware resources and the execution time required by different tasks, which makes it inapplicable to the real situation. The device placement for TensorFlow computational graphs is optimized using RL in \cite{P2}, which specifies device placement for each certain operation in the neural network. However, its scheduling can only generate a set of scheduling strategies for fixed tasks simultaneously. In DRL-Cloud \cite{DRL-Cloud}, it minimizes energy cost for large-scale cloud service providers with large number of servers by RL. In \cite{zx}, it presents a framework for power-efficient resource allocation in cloud radio access networks by RL as well.

\vspace{-0.5em}
\section{Conclusion}
By exploring the variability of workloads and performance requirements in driving automation and the heterogeneity of multi-accelerators, we purpose a comprehensive framework that synergistically handles the heterogeneous hardware accelerator design principles, system design criteria, and task scheduling mechanism. First, based on a taxonomy for emerging CNN accelerators, we design a heterogeneous multicore AI platform (HMAI) which adopts three typical CNN accelerator architectures. Next, by designing two metrics: Matching Score and Global State Value, autonomous driving system can guide the task execution on the platform. Finally, FlexAI-a reinforcement learning-based mechanism are proposed to generate scheduling policies in autonomous driving.

%\fi

%\input{file/test}

%%%%%%% -- PAPER CONTENT ENDS -- %%%%%%%%

%%%%%%%%% -- BIB STYLE AND FILE -- %%%%%%%%


\bibliographystyle{IEEEtran}
\bibliography{refs}
%%%%%%%%%%%%%%%%%%%%%%%%%%%%%%%%%%%%

\end{document}